# Putting high-index Cu on the map for high-yield, dry-transferred CVD graphene

*Oliver J. Burton[1‡*], Zachary C. M. Winter[2‡]*, Kenji Watanabe[3], Takashi Taniguchi[4], Bernd Beschoten[2], Christoph Stampfer[2,5], Stephan Hofmann[1]*

[1]Department of Engineering, University of Cambridge, Cambridge CB3 0FA, United Kingdom

[2]2nd Institute of Physics A and JARA-FIT, RWTH Aachen University, 52074 Aachen, Germany

[3]Research Center for Functional Materials, National Institute for Materials Science, 1-1 Namiki Tsukuba, Ibaraki 305-0044, Japan

[4]International Center for Materials Nanoarchitectonics, National Institute for Materials Science, 1-1 Namiki Tsukuba, Ibaraki 305-0044, Japan

[5]Peter Grünberg Institute (PGI-9), Forschungszentrum Jülich, 52425 Jülich, Germany

KEYWORDS: CVD, Graphene, Single Crystal, Dry Transfer, Data science, 2D material, High electron mobility

## ABSTRACT

Reliable, clean transfer and interfacing of 2D material layers is technologically as important as their growth. Bringing both together remains a challenge due to the vast, interconnected parameter space. We introduce a fast-screening descriptor approach to demonstrate holistic data-driven optimization across the entirety of process steps for the graphene-Cu model system. We map the crystallographic dependencies of graphene chemical vapor deposition, interfacial Cu oxidation to decouple graphene, and its dry delamination across inverse pole figures. Their overlay enables us to identify hitherto unexplored (168) higher index Cu orientations as overall optimal. We show the effective preparation of such Cu orientations via epitaxial close-space sublimation and achieve mechanical transfer with very high yield (>95%) and quality of graphene domains, with room-temperature electron mobilities in the range of 40000 $cm^2/(Vs)$. Our approach is readily adaptable to other descriptors and 2D materials systems, and we discuss the opportunities of such holistic optimization.

2D materials (2DMs), spearheaded by graphene, continue to be an extremely powerful platform for scientific discovery of ever more complex properties and functionalities. There is however a widening gap between individual demonstrator or "hero" devices and what is possible to reproducibly fabricate with scalable methodologies. This presents a key bottleneck for translation to technology, in particular for higher-value added applications such as integrated sensors, flexible high frequency electronics or broadband opto-electronics, as highlighted across current technology roadmaps.[1–3] Chemical vapor deposition (CVD) has matured as the leading technique to scalable crystal growth of mono-/few-layer graphene[4–6], and as-grown synthetic material has reached the quality (as defined by electron mobility measurements) set by exfoliation from bulk crystals.[7–9] Most applications involve transfer away from the growth substrate, and such transfer and handling technology is thus an integral part of the scalable CVD approach.[3,10] Given the notoriously vast, combined parameter space, to date graphene CVD and transfer optimization has largely been explored in separation, with all early focus on the initial synthesis parameters and utilizing catalytic enhancement via transition metals such as Cu.[11,12] Such catalytic growth of graphene has a high dependence on Cu facet orientation, whereby most recent growth studies converged on using Cu(111),[13–16] owing to ease of production of such low index orientation both via foil crystallization and epitaxial metallization approaches, as well as enabling a uniform epitaxial alignment of graphene. In order to promote transferability, the graphene-Cu interaction must be weakened post-growth, to decrease graphene adhesion enough for clean, reliable delamination and for the Cu template to be re-used. An efficient approach for this is interfacial Cu oxidation[8,17,18]. Such post-growth process is also known to have a high dependence on the Cu surface orientation.[19] A common observation across many different oxidation approaches is the low achievable rate of oxidation of Cu(111) underneath graphene.[19–22] This indicates the shortcomings of the current sequential optimization approach, where graphene on the growth substrate might be "high quality" but subsequent transfer is compromised, and so will be device yield and achievable properties.

Here, we use a fast-screening descriptor approach to demonstrate a holistic, combined optimization approach across the entirety of process steps for growth and transfer for the graphene-Cu model system. We focus on enabling efficient dry-transfer of CVD graphene islands, as this is currently a much sought after capability and a critical first step to address the demand for reproducible, high yield device fabrication relying on cleanly interfaced 2D material stacks. We systematically track and study 1000's of graphene islands on over 100 crystallographic Cu orientations, and plot quality descriptors for each process step across inverse pole figures (IPF). This representation allows us to overlay IPFs to identify higher-index Cu orientations that are best suited for the combined overall process. We employ an epitaxial close-space sublimation approach[15] to exclusively create optimum (168) Cu orientation, establishing translation to a scalable pathway for graphene island growth and transfer at high (>95%) yield. After h-BN encapsulation, we demonstrate room temperature electron mobilities of over $40\times10^{3}$ $cm^{2}/Vs$ at $1\times10^{12}$ $cm^{-2}$ and average Raman 2D line widths of ~16 $cm^{-1}$. We find this approach extremely powerful to navigate and gain new insights across these notoriously large, interconnected parameter spaces, and readily adaptable to many other catalyst-2D material systems whilst being expandable to include future relevant descriptors.

## RESULTS AND DISCUSSION

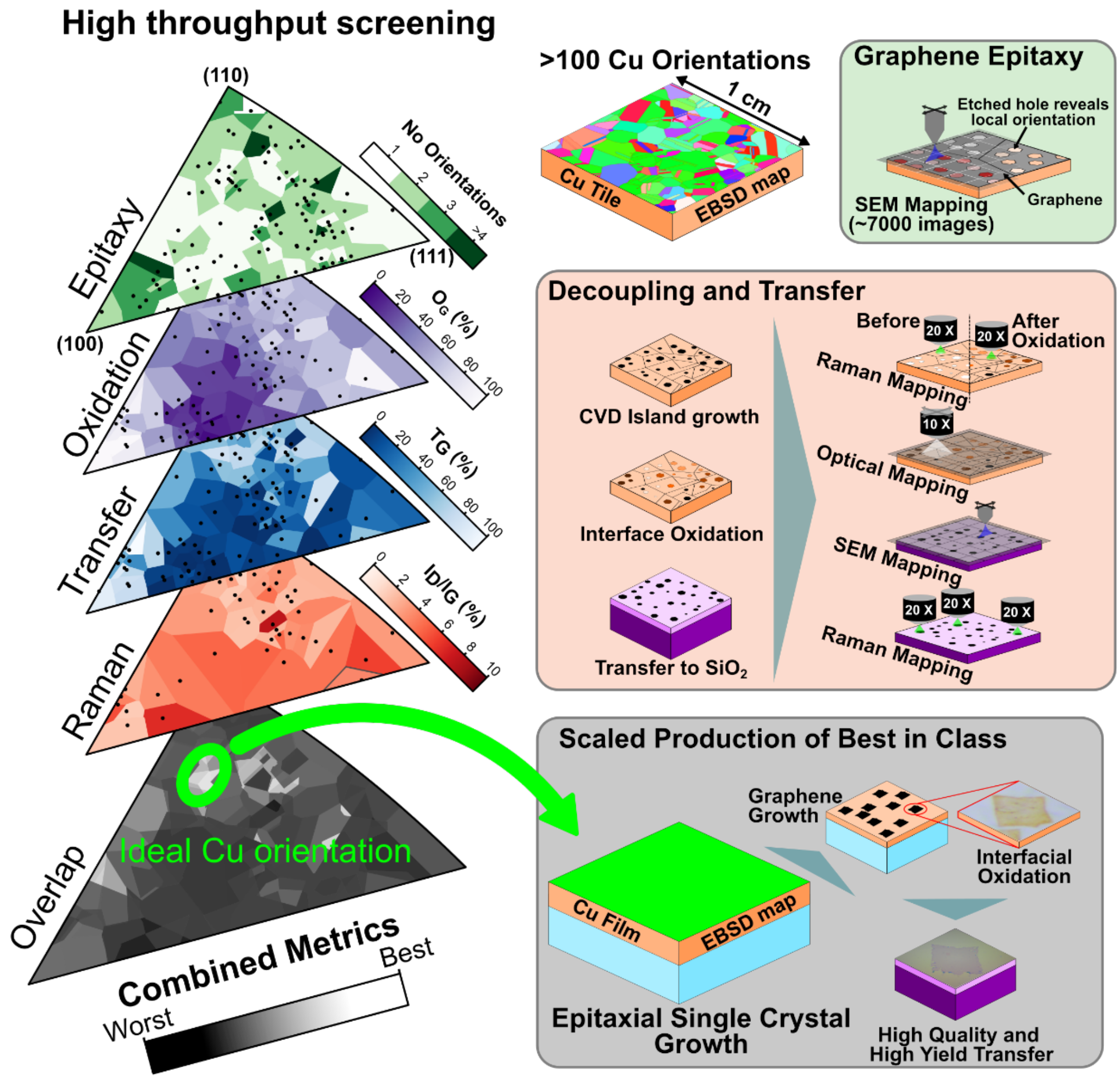

Figure 1: Schematic summary of the holistic, high-throughput characterization methodology and epitaxial production of select orientations. Each key screening parameter is mapped on an inverse pole figure representing catalyst crystallographic orientation, with the combined metrics (an overlay of these screening parameters) revealing the optimum Cu orientation. These metrics include the graphene alignment, interfacial oxidation propensity, transferred percentage and Raman characterization for individual graphene islands. The optimum Cu orientation is then reproducibly created using an epitaxial growth platform, enabling high yield graphene island transfer.

We utilize two different catalyst preparation methods, as outlined in Figure 1. First, the use of polycrystalline, 1x1 $cm^2$ Cu tiles (See Methods) exhibiting a large number (>100) of different Cu crystallographic orientations, each of them sufficiently large (>100μm), allowing for the effective high-throughput characterization of graphene growth, interfacial Cu oxidation to decouple graphene, and its mechanical delamination. We fully map the surface crystal orientations of the Cu tiles by electron back scatter diffraction (EBSD). For each process step we identify a key quality metric (Figure 1; each of which is discussed in detail below) that can be effectively, automatically mapped and compiled as an IPF. By overlaying individual process step IPFs, the use of polycrystalline Cu tiles thus allows the identification of Cu orientations that are overall most promising throughout the growth and transfer parameter space. To selectively work with as-identified optimum Cu orientations we employ an epitaxial close-space sublimation approach[15] as a second catalyst preparation method that enables the scalable production of single crystal metal templates. We use graphene islands grown on these single crystal templates to characterize the graphene in terms of the reproducible transfer yield of multiple islands, and through the fabrication of encapsulated test devices to confirm a high-quality material.

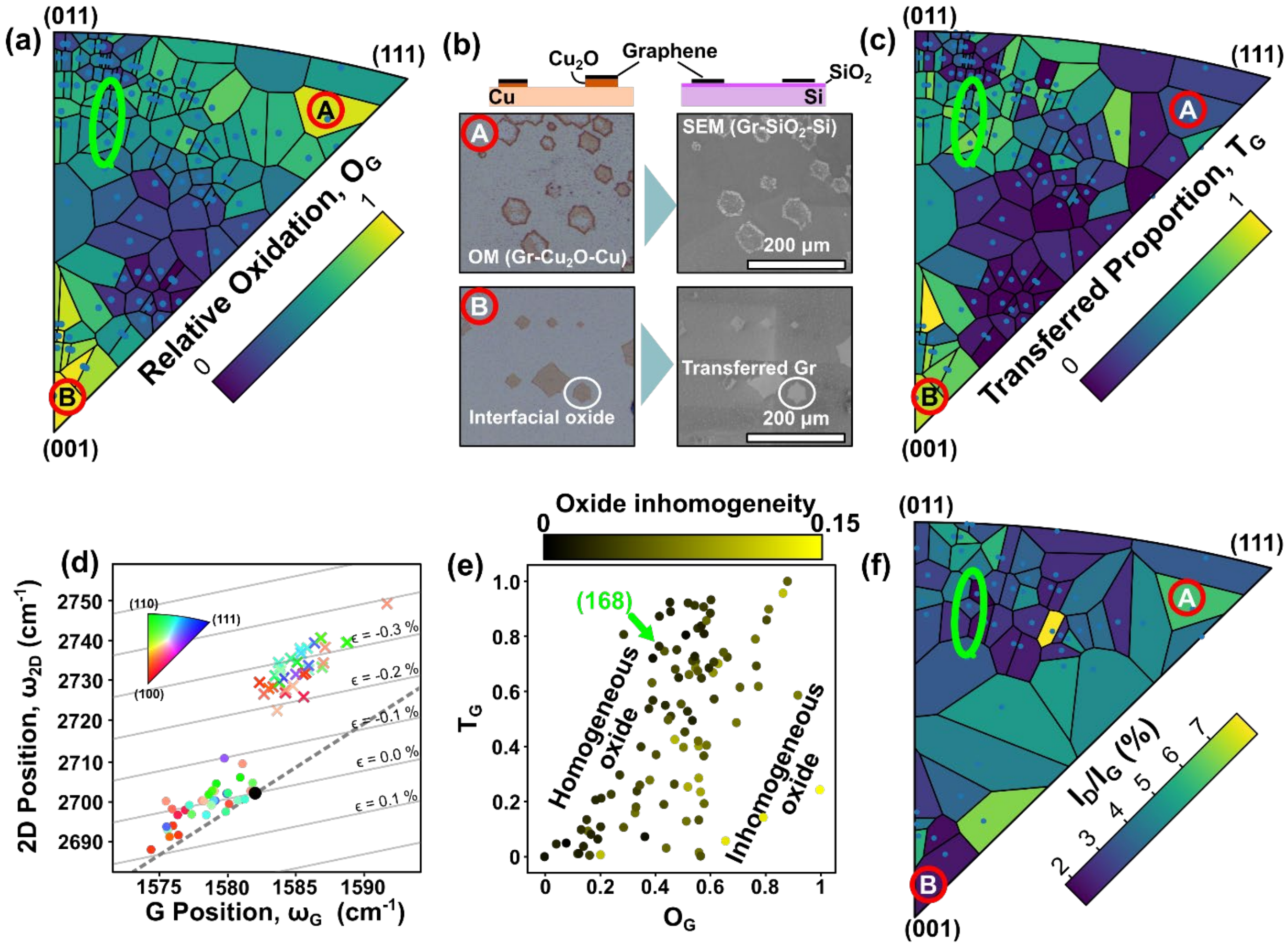

Figure 2: The mapping of interfacial oxidation, graphene transfer yield, and quality after transfer onto SiO2-Si as defined by Raman $I_D/I_G$ peak ratios. (a) A IPF showing the relative oxidation of the Cu-Gr interface, $O_G$, dependent on crystallographic Cu orientation. (b) A schematic illustration of the ordering of the Gr-$Cu_2O$-Cu and Gr-$SiO_2$-Si samples with example images from OM and SEM shown below, labeled (A) and (B) with corresponding identifiers on the IPFs. (c) A IPF showing the fraction of transferred graphene, $T_G$, from Cu after oxidation as a function of Cu orientation. (d) The mean Raman 2D peak position against the G peak position for graphene on Cu before (x) and after (o) oxidation with lines of constant strain labelled and a line of constant doping shown for reference and the pristine point

(the point with no strain or doping) is shown as a large black point. The color of this scatter plot is linked to the crystal orientation of the Cu that the graphene measurement was taken, according to the IPF inset. (e) A scatter plot of $T_G$ vs. the relative oxidation, $O_G$, with the color of each data point mapping to the inhomogeneity of the relative oxidation (See Methods). (f) a IPF of the $I_D/I_G$ Raman peak ratio of Gr after mechanical delamination and transfer onto a $SiO_2$ substrate as a function of Cu orientation. The blue dots in the IPFs in this figure represent the average crystallographic orientation of the Cu facet as measured by EBSD, and the green ellipse represents the ideal region around Cu(168).

In this work, we focus on the mechanical delamination of graphene from its growth substrate. To generate a significant number of data points for analysis, we focus on individually grown graphene islands on a polycrystalline Cu catalyst. Motivated by prior work, we use saturated water vapor exposure to promote interfacial Cu oxidation[8,19,23] prior to delamination with a PVA film (See Methods). Figure 2 connects the relative oxidation level beneath graphene islands to the yield of their mechanical delamination and quality of as-transferred graphene on $SiO_2$ support. The clear optical contrast due to Cu oxidation allows us to employ optical microscopy (OM) and introduce a quantitative parameter $O_G$. Here, $O_G$ represents the normalized mean relative Cu oxidation contrast of the areas beneath all graphene islands on a given Cu orientation (See Interfacial Oxidation in Methods). Figure 2(a) shows an IPF for $O_G$ and demonstrates the variation in interfacial oxidation as a function of Cu orientation. The interfacial oxidation was further characterized by X-ray photoelectron spectroscopy (XPS) and imaging ellipsometry (IE) to confirm that the high throughput screening via $O_G$ is indeed a meaningful metric (See SI Figures S5 and S7). To capture transfer yield, we use scanning electron microscopy (SEM) mapping and introduce a quantitative parameter $T_G$ (See Mechanical Delamination in Methods) that reflects the average proportion of graphene islands that are transferred onto $SiO_2$ via the mechanical delamination process. Figure 2(c) presents $T_G$ from the Cu growth substrate to a $Si/SiO_2$ substrate as an IPF.

Figure 2(e) presents a scatter plot of $T_G$ vs $O_G$ to highlight the relationship between these two parameters. The data points are colored to highlight oxide inhomogeneity (See Interfacial Oxidation in Methods), a measure of the variation of the interfacial oxide from the mean with lower values meaning that the oxide is more homogeneous (*i.e.,* exhibiting a more uniform contrast). Our data identifies a clear underlying trend: the more interfacial Cu oxide, and the more homogeneous that oxide, the higher the proportion of successfully transferred graphene. It is notable that our measured relative oxide thicknesses and $O_G$ trends across the IPF are consistent with previous literature[19], despite different oxidation and exposure conditions. This indicates that the trends we show are representative across a reasonably large set of potential oxidation conditions. Example OM and SEM images are labelled A and B (approximately Cu(111) and Cu(100) respectively) in Figure 2(b), with corresponding locations marked on Figures 2(a,c,f), highlight how more oxidized Cu regions link to a higher success rate for graphene transfer. It is noted in example data (A) that whilst a visible degree of oxidation can be seen beneath the center of some graphene domains, these areas are not successfully transferred, whilst the more oxidized regions at the edge of the graphene islands are. This implies one, or a combination, of three scenarios: (1) that there is a threshold oxide thickness requirement for decoupling and delamination, (2) that the oxide inhomogeneity is higher spatial frequency than the resolution of the imaging techniques used or, (3) that the oxide in the center of these graphene domains is different to that of the outside, *i.e.* $Cu_2O$ vs. CuO, and couples strongly to the graphene. We can rule out the latter, as XPS on Cu(111), similar to the orientation in question, and on all other Cu facets measured (See SI Figure S7) shows a lack of $Cu^{2+}$ at the surface. This implies that the dominating oxide formed at the Cu-graphene interface is $Cu_2O$. Whilst example areas A and B demonstrate the importance of the oxide homogeneity and presence, they also highlight the complexity of the system. There are different oxidation mechanisms that have significant variations in both rate and propensity of lateral oxide propagation. Example A in Figure 2 demonstrates this lack of propensity with only thick oxide observed at the edge of the graphene islands. Single crystal prepared Cu(111) shows this lack of oxidation as well, with our tests (similar graphene islands on single crystal Cu(111) in the same humidified environment described in methods) showing no propagation of oxidation beneath graphene grown on Cu(111) even after several weeks of oxidation. This lack of oxidation is consistent with prior literature, which suggests that Cu(111) inhibits extended oxidation beneath graphene[20] and speculates that this links to the close commensurate matching and thus coupling of in-plane graphene and Cu(111).[24] However, by examining the Cu regions not covered by graphene (See SI Figure S13) we reveal that there is in general a strong correlation (0.770 Pearson correlation coefficient) between the oxidation of Cu facets beneath the graphene and of the uncovered Cu. This strong correlation implies that most of the variation in oxidation between facets under graphene is also seen on bare Cu, so decreases in propensity to oxidation is unlikely to be the result of graphene.

Figure 2(f) shows the graphene D to G Raman peak intensity ratio ($I_D/I_G$) as a function of Cu orientation as an IPF plot. The $I_D/I_G$ ratio is a commonly used metric in the literature with higher values corresponding to a higher defect density in graphene.[25] Figure 2(f) shows similar tendencies as Figure 2(c): Areas with lower $T_G$ have a higher $I_D/I_G$, or statistically speaking the $I_D/I_G$ vs $T_G$ has a Pearson correlation coefficient of -0.38, implying that Cu orientations with a high $T_G$ tend to yield a graphene film with lower defect densities after transfer. We postulate the variation of $I_D/I_G$ with Cu orientation corresponds to defects in the film as a result of cracks and holes formed through the graphene transfer process, rather than any intrinsic variation in the quality of the graphene as grown on different Cu orientations. This cracking and its effects on Raman spectroscopy measurements can be seen in SI Figure S3(d). Figure 2(d) plots 2D Raman peak position $\omega_{2D}$ against $\omega_G$ for a range of Cu orientations, showing a strong difference between the Raman peak positions, which have been correlated to strain,[26] before and after oxidation. These peak shifts imply that as-grown graphene on all measured bare Cu facets is under compressive strain, which upon interfacial Cu oxidation reduces or shifts to tensile strain.[20] This shift is consistent with the volume expansion upon Cu oxidation,[27] given a Pilling-Bedworth ratio of 1.7 for $Cu_2O$. It is noted that we have adjusted the pristine point (*i.e.*, the value of $\omega_{2D}$ and $\omega_G$ representing no strain or doping) for the laser wavelength used (457 nm) according to the Raman peak dispersion experimentally determined in the literature.[28,29]

An analysis of the 2D and G Raman peak widths (Γ) shows that initially there is a wide range of $\Gamma_{2D}$ and $\Gamma_G$ before oxidation, narrowing after oxidation to a much smaller region of higher average $\Gamma_{2D}$ and $\Gamma_G$ (See SI Figure S3(a)). Literature has previously established different $\Gamma_{2D}$ for graphene on different crystallographic orientations of Cu, which is notably reflected within SI Figure S3(a) with Cu(111) having a broader 2D band than both Cu(110) and Cu(100).[30] Previous experiments have linked the increase in $\Gamma_{2D}$ to an increase in the magnitude of nanoscale strain variations.[31] This correlates well with our measurements of the surface microstructure of as-oxidized Cu facets beneath graphene layers, which are microscopically rougher than the initial metallic Cu at the interface (example in SI Figure S9), consistent with reports across the literature.[32,33] The shift to lower Raman peak positions and increase in widths implies that the graphene is moving from a region of consistent compressive strain to a region of tensile strain with larger variations in local strain. We interpret this as the graphene being detached from its relatively strong coupling to the bare Cu to rest on a rougher Cu oxide surface, which then facilitates mechanical delamination. The $\Gamma_{2D}$ of the graphene on Cu here show significantly higher values than after transfer onto the $SiO_2$-Si substrate (See Figure 2(c)). However, we observe no clear dependency on Cu surface orientation between the $\Gamma_{2D}$ before and after mechanical delamination, implying that the $\Gamma_{2D}$ measured before transfer is a poor predictor or quality metric of any graphene characteristics after transfer onto another substrate. The data shown in Figure 2 shows that for the graphene-Cu system reproducible mechanical graphene delamination with a low defect density requires effective full and homogeneous oxidation of the buried Cu interface to graphene.

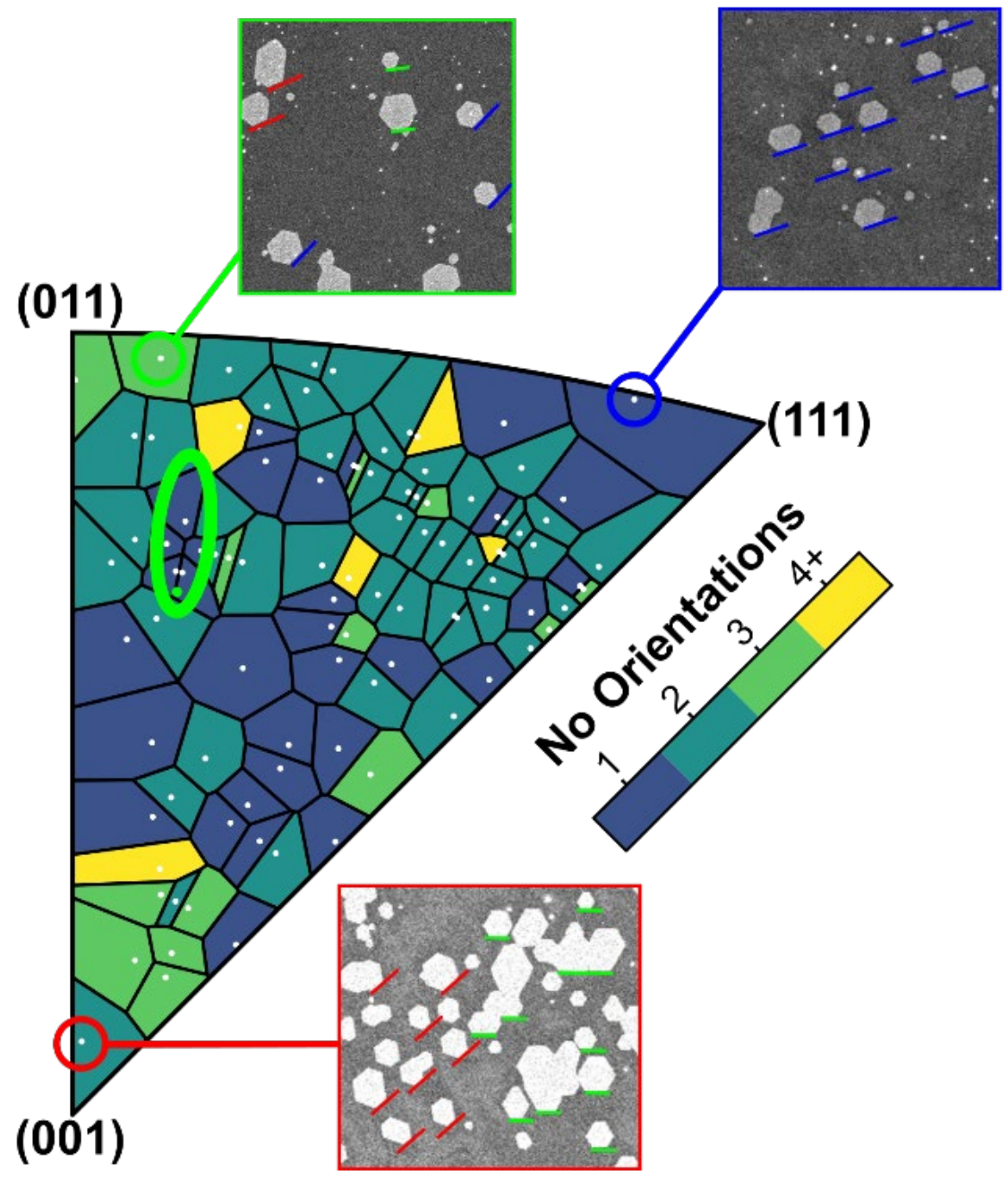

Figure 3: A IPF showing the number of graphene orientations grown on each measured Cu orientation (white points correspond to mean EBSD measurement for that facet). The SEM images show etched holes in graphene on select Cu orientations indicated on the IPF, the number of orientations of graphene etch holes found on each Cu facet is indicated by the color bar. The green ellipse highlights the region around Cu(168).

Given the strong intercalation and growth dependence on crystallographic alignment, the characterization of different potential graphene island orientations is important. We use a post-growth Ar/$H_2$ gas mixture after the graphene growth process to etch small hexagonal holes into a graphene film, locally exposing zig-zag edges of graphene.[15] Analyzing their orientation allows an effective mapping of the local crystallographic orientation of the graphene. By combining 6515 SEM images, detecting and measuring the orientation of these zig-zag edged holes and determining their location (383 676 found holes) across the tiled Cu sample we compiled Figure 3: an IPF map of the number of orientations of graphene found on each Cu facet orientation (See SI Figure S4 for additional details). Given our findings of Figure 2, it is highly preferred for the CVD process to lead to a single, uniform graphene island alignment. This has been shown to be also a pre-requisite for single crystalline graphene films,[34–37] and most literature has thus focused on Cu(111). Our data for the low index Cu orientations is consistent with prior literature: there are three orientations of graphene on ~Cu(110),[38] two orientations of graphene on Cu facets tending towards (100),[35,36] and a single orientation of graphene grown on Cu(111).[35,37] Figure 3 shows though that there are a number of higher index Cu orientations which also give a single graphene orientation. A direct overlay with IPFs in Figure 2, (shown in Figure 1) particularly motivates the cluster of higher index orientations around Cu(168). CVD graphene on these facets shows not only a single epitaxial orientation but amongst the lowest $I_D/I_G$ ratios and $\Gamma_{2D}$ widths (See SI Figure S6 for facet selection information), as well as high $O_G$ and high $T_G$.

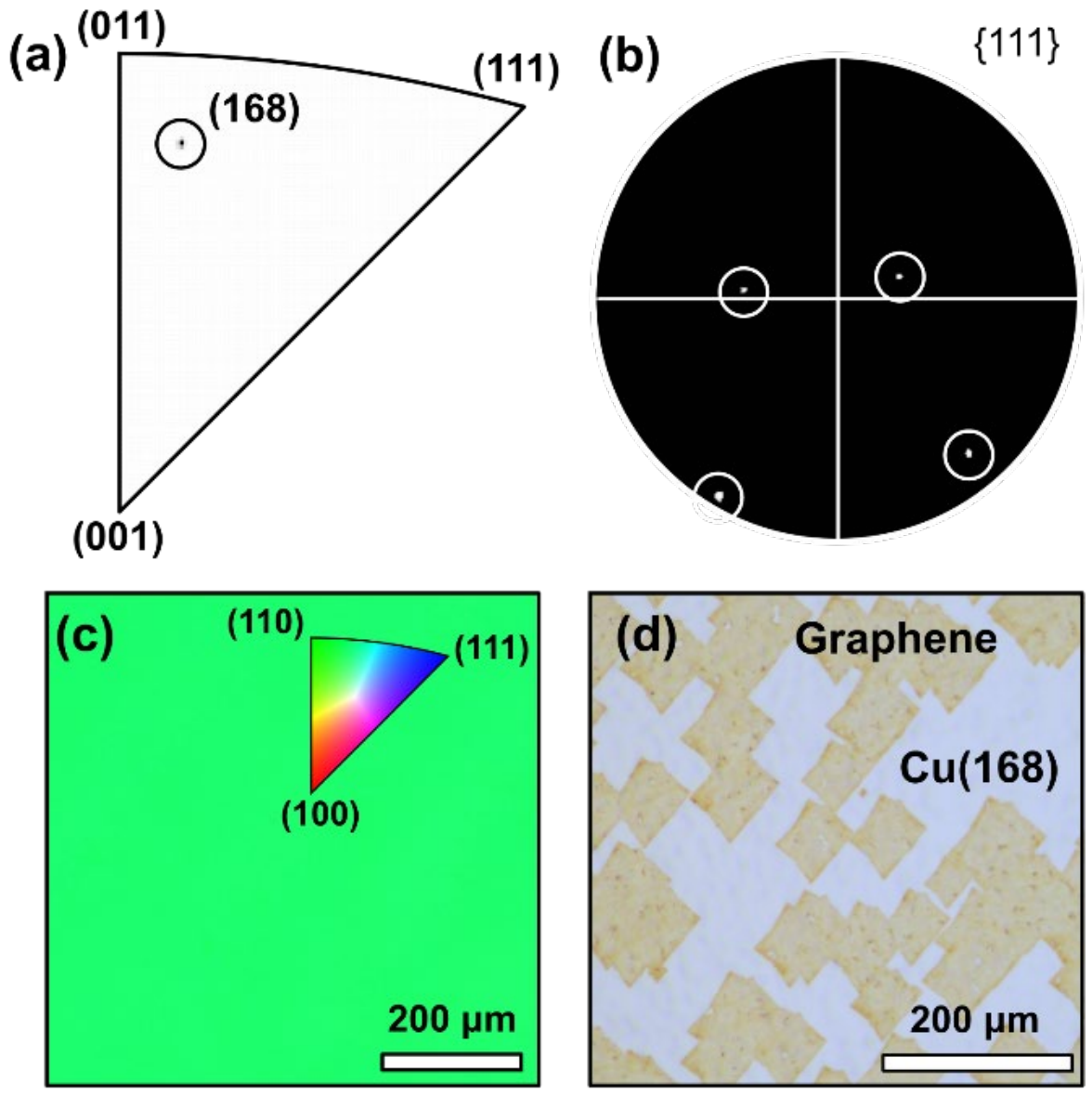

Figure 4: The characterization of epitaxial closed-spaced sublimation Cu(168) film deposition on MgO(168) substrates. (a) A density IPF of the spatially mapped EBSD data. (b) A {111} pole figure of the EBSD data. (c) The IPF colored spatial map, inset showing map of color to crystallographic orientation. (d) An OM image showing the graphene islands (orange) after the interfacial oxidation.

Having identified orientations around Cu(168) as the optimum surface for growth and transfer using polycrystalline Cu tiles, Figure 4 shows that such Cu orientations can be selectively prepared via epitaxial Cu growth on MgO substrates. We employ an epitaxial closed spaced sublimation (CSS) approach that we previously introduced for single crystal Cu(111) wafer growth (See Methods).[15] This allows scalable, cost-efficient epitaxial metallization at comparatively high rates and can be seamlessly combined with the graphene CVD process. Figure 4 (a)-(c) show results of EBSD mapping and analysis of approximately 10 μm thick CSS Cu(168) films on cm-sized MgO(168). The IPF shows the creation of Cu(168), the {111} pole figure demonstrates that there is only one in-plane orientation of Cu(168) and the spatially resolved IPF map shows that this is consistent over large areas; these all confirm that the Cu films are single crystal over the analyzed ~$1\times1$ $mm^2$ region. The absence of any thermal grooving observed by OM is further consistent with the single crystallographic nature of as-grown epitaxial Cu. Figure 4(d) shows a representative OM image of graphene islands grown on such epitaxial Cu(168) after oxidation. Consistent with Figure 3, we observe a single graphene island orientation. Consistent with Figure 2, we observe homogeneous interfacial oxidation.

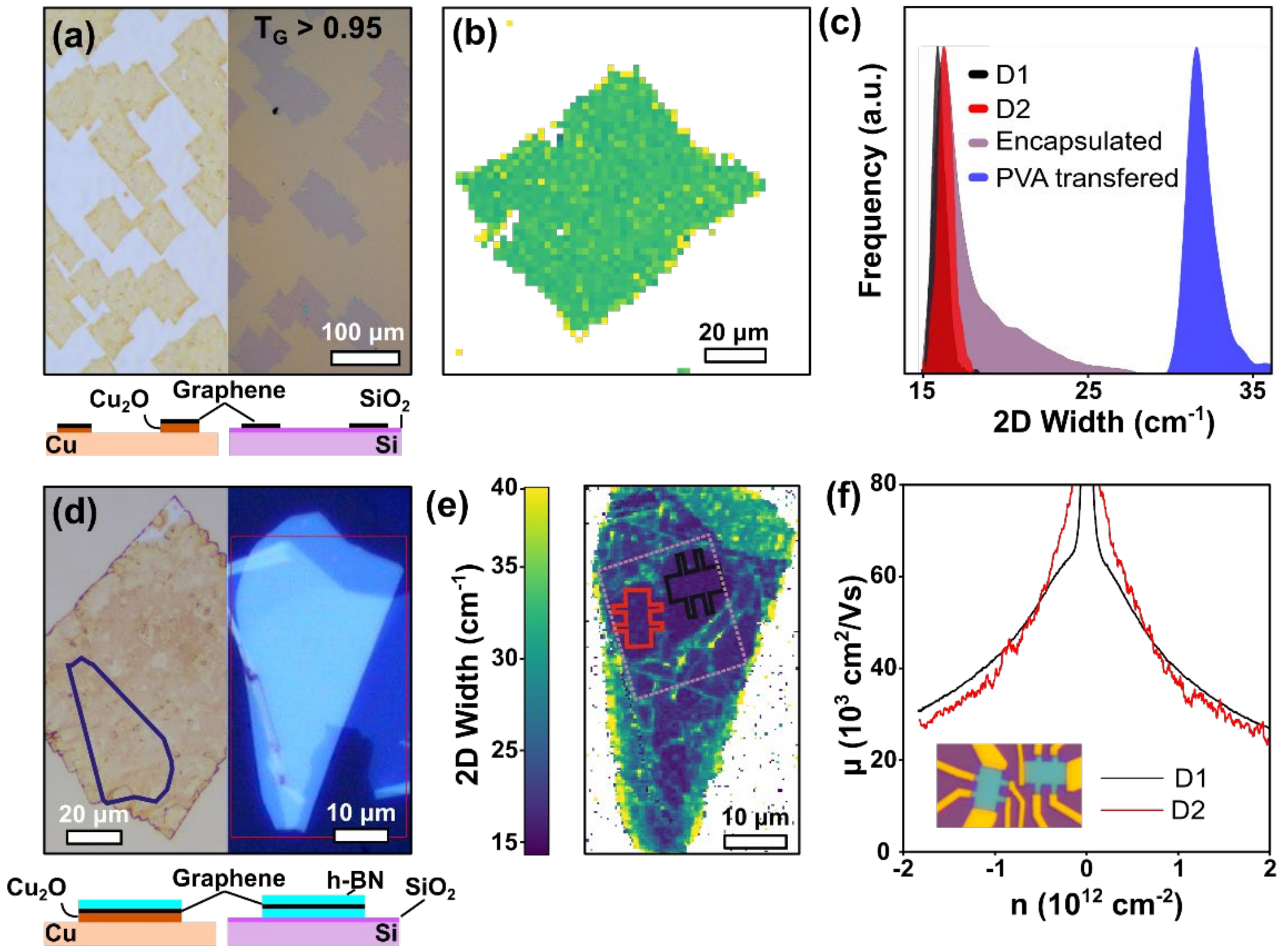

Figure 5: Yield and quality measurements of CVD graphene grown on CSS deposited Cu(168). (a) OM images of the graphene islands before (left) and after (right) transfer with PVA onto Si/$SiO_2$, the lower schematic shows the ordering of materials from a side view. (b) A spatially resolved map of the Raman 2D peak width of the graphene on Si/$SiO_2$ shown in (a), with the colormap the same as that in panel (e). (c) A histogram of Raman 2D peak width of the data in (b) shown in blue, and the data from (e) shown in black, red, and purple corresponding to the signal from the shapes outlined in (e). (d) OM images before and after encapsulation of the graphene in h-BN flakes, with schematic images showing ordering below. (e) The Raman 2D width map of the encapsulated region shown in (d), with regions used to make devices outlined. (f) The mobility as a function of carrier concentration at room temperature for both devices, with line color corresponding to the respective regions in (e).

After finding the ideal growth-oxidation-transfer path through intersecting the respective parameter spaces, we use two simultaneous approaches to probe (1) the reproducible yield and (2) the quality of the graphene islands from Cu(168): (1) a PVA style transfer identical to that used above for the mechanical delamination from polycrystalline Cu tiles; (2) mechanical dry-delamination of graphene with exfoliated h-BN crystals to fabricate encapsulated Hall bar devices. This combined approach allows us to quantify process reproducibility in terms of $T_G$ and quality in terms of Raman spectroscopy and Hall mobilities. Figure 5(a) shows the graphene after oxidation on epitaxial CSS Cu(168), and after transfer on $Si/SiO_2$. The optical contrast for the former highlights full and homogenous Cu oxidation underneath the graphene islands. The analysis of graphene islands over 3x3mm area of the single crystal Cu(168) shows a $T_G > 0.95$. For comparison we carried out the same process and analysis for epitaxial Cu(111), Cu(123) and Cu(120) (See SI Figure S14), which shows a $T_G < 0.1$ for Cu(111) and Cu(123) and $T_G < 0.5$ for Cu(120). This is consistent with Figure 2, *i.e.*, our results from the polycrystalline Cu tile screening, and highlights the achieved yield increase in transfer. We note that even after long (4 weeks) oxidation, epitaxial Cu(111) did not oxidize fully, consistent with previous research.[19] This implies that for graphene grown on Cu(111), harsher and more damaging oxidation treatments are required to oxidize to the same standard as on Cu(168). This underscores our argument of the superior yield/quality balance achievable for such higher index Cu orientations. In order to highlight achievable graphene quality, we use Hall bar devices based on widely used h-BN heterostructure encapsulation, to avoid well-known substrate, particularly $SiO_2$, dependent scattering effects,[39] and to allow direct measurement of mobility. For this, we follow previous literature[8,9], using a stamp terminated with a h-BN flake to cleanly transfer the graphene from oxidized CSS Cu(168) (see methods and SI Figures S8 & S12 for fabrication details). Figure 5(d) shows both the graphene island on the Cu(168)/$CuO_2$ and the h-BN flake on Si/$SiO_2$ used for subsequent device fabrication. Figure 5(b) shows a Raman 2D peak width map representative of an unencapsulated graphene island transferred on Si/$SiO_2$. The Raman 2D peak width $\Gamma_{2D}$ is empirically and theoretically linked to both the quality of the graphene and the effects of its support and interfacing,[40,41] including nanometer-scale strain variations.[31] Combined with the OM data of Figure 5(a) the map highlights consistent quality across the graphene area. Figure 5(e) shows a map of $\Gamma_{2D}$ of graphene encapsulated in h-BN. Some folds and bubbles are seen, typical of this style of fabrication. Outlined in black and red in Figure 5(e) are two Hall bar device footprints in the most homogenous regions, which show a mean $\Gamma_{2D}$ of 16 $cm^{-1}$. Figure 5(c) compares the Raman $\Gamma_{2D}$ between graphene on $SiO_2$ and when encapsulated in h-BN . These $\Gamma_{2D}$ histograms shows that the h-BN encapsulated graphene has significantly lower 2D peak widths than the graphene on Si/$SiO_2$ (with mean $\Gamma_{2D}$ of 32 $cm^{-1}$ and 16 $cm^{-1}$ respectively), and reflects typical values of exfoliated graphene found in the literature[40] and other state of the art pick-up techniques.[7,8] The ability to mechanically directly delaminate graphene with a h-BN stamp further highlights the effective decoupling of the graphene via interfacial Cu oxidation, that is, the $Cu_2O$ decreases the adhesion of the graphene to the growth substrate to below that of a graphene/h-BN interface, not requiring an intermediate wet transfer such as in other state-of-the art encapsulation techniques[7]. Figure 5(f) shows the measured charge carrier mobility, μ, as a function of carrier concentration at room temperature. The devices show consistent performance with a mobility of $42.1\times10^3$ $cm^2/Vs$ at $1\times10^{12}$ $cm^{-2}$, indicating a quality at par with state-of-the-art exfoliated graphene and previously reported best results for CVD graphene.[7,8]

## CONCLUDING REMARKS

High quality graphene grown on substrates that are incompatible with further processing techniques does not answer the question of how to bring the promised performance of graphene to industry and high value-added applications. Here we have applied a holistic, data-driven approach to process optimization, utilizing fast-screening descriptors across the entirety of process steps for growth and transfer for the graphene-Cu model system. Our IPF overlay methodology allowed us to identify clear advantages of hitherto unexplored higher index Cu orientations. The increase in yield for the dry transfer of isolated CVD graphene islands shown here is essential for the many ongoing efforts to automate[42] and accelerate device assembly that relies on heterostructures of increasing complexity, including stacking angle or designer (meta) materials. Our approach is readily adaptable to many other catalyst-2D material systems, *e.g.* $WS_2$/Au,[43] h-BN/Cu,[44] or h-BN/Pt[45] that are held back by analogous challenges. We anticipate the introduced high-throughput IPF based methodology to become a potent platform to explore the many hitherto not well understood orientation dependencies of chemical reactions and physical effects confined between a 2D layer and metal/substrate[46,47], as well as emerging epitaxial systems[43–45,48] across many related material systems, with additional relevant descriptors easily added.

## EXPERIMENTAL DETAILS

**Plotting:** The IPFs shown in this work contain regions, colored according to a provided color map, which correspond to the associated datapoint's Voronoi cell where all points in that cell are closer to the contained datapoint than any other. This is done in absence of a continuum of data to clearly present regions of interest and represent the computational methods described elsewhere. It is noted that large cells are not necessarily representative of the areas they cover, and for clarity the position of each data point is clearly indicated in each cell in all IPFs in this work.

**Graphene growth** was done using previously defined CVD parameters[49], consisting of oxidizing the Cu surface at 200 °C for 30 minutes, heating up in BM Pro 4" CVD reactor (base pressure 4 x $10^{-2}$ mbar) to approximately 1065 °C where it is kept for all processes, annealing in Ar (650 sccm; 50 mbar) for 30 min, annealing in $H_2$ and Ar (100:500 sccm; 50 mbar) for 60 min, followed by Ar, $H_2$ and $CH_4$ (0.32:64:576 sccm; 50 mbar) for 5 min to grow graphene islands. The reactor was cooled down at base pressure with no gas flow.

**Graphene orientation mapping** was carried out on continuous graphene (aforementioned gas ratios, growth time extended to 1 hour), where the sample was then exposed to $H_2$ and Ar (170:470 sccm; 50 mbar) immediately after growth for 20 min. This yielded small (~5-10 μm diameter) holes with a hexagonal shape. SEM was then used to spatially map all holes over the Cu tile: approximately 7000 SEM images at 1024x786 resolution at 600x magnification. These images were then binarized, stitched and processed as detailed in Figure S4 to measure the angle of the etched hole, which was then linked spatially to the EBSD map to bin these measurements into Cu orientations. The orientations of graphene in each Cu orientation were then processed into a frequency density plot and the number or orientations was dictated by the number of peaks found by the SciPy python package's 'find_peaks' function.

**Mechanical delamination** was done using PVA for the systematic 'tile' studies: 7g PVA (8000-10000 MW, 80% hydrolysed; Sigma Aldrich) and 3g PVA (85000-124000 MW, 87-89 % hydrolysed; Sigma Aldrich) was mixed with 40 mL DI water and stirred at 80 °C until fully dissolved. Approximately 0.1 ml $cm^{-2}$ was placed on a removable support and dried at room temperature in a cleanroom environment. SI Figure S8 outlines the peeling process: The PVA film was then placed onto the dried graphene/Cu/sapphire at 120 °C to soften the PVA film allowing it to adhere to the graphene and conform to the surface. The PVA/graphene was removed from the Cu at room temperature and placed onto the Si/$SiO_2$ at 120 °C and left for 1 min. Once the PVA was cool, the PVA/graphene/substrate was placed in DI water at 80 °C for > 24 h to dissolve the PVA. The fraction of graphene transferred for each crystallographic orientation, $T_G$, is calculated by summing the areas of graphene after transfer, and taking this as a ratio with those area that contained graphene prior to transfer:

$$T_{G,hkl} = \frac{\sum(area\ of\ graphene\ islands\ before\ transfer)}{\sum(area\ of\ graphene\ remaining\ in\ islands\ after\ transfer)}$$

Areas were calculated by counting the number of pixels that contained graphene, and scaling by the spatial dimension of each pixel.

**Dry-transfer with h-BN** was done as shown in previous literature[9] using a stamp consisting of 13% PVA and 50 K PMMA. The polymers were spin coated onto a glass slide at 1000 RPM and heated at 110 °C for 10 min. Subsequently, h-BN was exfoliated first using Minitron 1008R tape multiple times to decrease the thickness of crystalline h-BN and then brought into contact with the stamp. A h-BN flake of appropriate thickness was located using confocal microscopy and a small area was cut. This stamp was then placed onto a Gel-pack polysiloxane based support layer, like PDMS, and glass slide. The glass/PDMS/PVA/PMMA/h-BN could then be brought into contact with graphene on the pre-oxidized copper substrate. After picking up the graphene, it was placed onto h-BN already exfoliated onto $SiO_2$.

**Raman spectroscopy** was done on Si/$SiO_2$ substrate using a Renishaw InVia system at 20x magnification using a 532 nm laser (at 10 % laser power) with counts accumulated over 1 s. The Si/$SiO_2$ substrate was leveled prior to measurement to ensure a consistent focus prior to the batch measurements across the 1 $cm^2$ sample. Each of the (~400) spectra in each of (127) maps were fitted using separate Lorentzian profiles for the D, G and 2D peaks. Spectra corresponding areas where there is little/no graphene were discarded before statistics were formulated to minimize noise, defined as spectra below a threshold number of counts (400 counts in this work). Data sets with less than 10 accepted spectra were discarded to ensure a reasonable sample size per map and prohibit noise from significantly influencing the results. The values shown in the main text correspond to the statistical mean values after fitting and filtering of data. For the Raman spectroscopy studies on graphene on Cu we used a Witec Raman Imaging Microscope alpha300R and the spectra were obtained with a 50 x objective, equipped with a x-y-z DC piezo stage with the positions manually correlated to the crystallographic orientation. For excitation a 457nm laser was used to limit the influence of the Cu background[50]. The Raman maps were sampled before and after oxidation as shown in Figure S10. The maps are further processed to average values which could be correlated with the Cu crystallographic orientation as shown in Figure S11.

**Optical microscopy** was used to stitch together images using a 10 x objective to create a final image of 16353 by 15752 pixel image. To remove the non-uniform contrast and brightness in the final image, a global background removal was used to remove the horizontal and vertical contrast profiles present in the stitched image that this non-uniformity caused:

$$P_{ij} = Q_{ij} - \frac{\sum_{i=0}^{N} Q_{ij}}{N} - \frac{\sum_{j=0}^{M} Q_{ij}}{M}$$

where $P_{ij}$ is the value of the new pixel in each color channel, $Q_{ij}$ is the original pixel and N and M are the height and width of the image in pixels respectively. This homogenized the image and provided a means to globally compare the contrast due to oxide of individual islands.

**Electron backscatter diffraction** (EBSD) maps were created with a FEI Nova NanoSEM instrument at 30 kV with a 500 µm aperture. The sample was tilted to 70° approximately 17 mm from the pole piece, with the EBSD detector screen approximately 20–25 mm from the sample. The EBSD was calibrated and optimized for Cu patterns to ensure a successful fit rate of close to 100%. The individual grains from these maps were then identified, and the (directional) mean Euler angles were used to create the larger stitched map used in this study. In this work EBSD was conducted on the Cu tile after the growth, oxidation and peeling of the graphene.

**Interfacial oxidation** was achieved using water vapour as shown in SI Figure S2. This consisted of heating water in the base of closed desiccator on a hotplate to 70 °C, translating to a sample temperature of approximately 30 °C on the sample stage with a measured humidity of ~99 %. An angled glass plate was placed over the Cu substrates to prevent any water condensate from contaminating the samples. The oxidation time used in this work was 4 days, apart from the direct on Cu RS data (Figure 2(d) and SI Figure S3(a)), where the time period was 2 days. This was chosen as it was observed that after this time the interfacial oxidation was no longer progressing at a noticeable rate on difficult to oxidize facets, and it can be seen that the oxide IPF shown in Figure 2 matches that of previous work[19] over much longer time frames. The relative extent of interfacial oxidation, in this work referred as $O_G$, is calculated based on the optical contrast extracted from OM images. This takes all pixels within all islands of graphene on a particular Cu orientation and takes the mean of these values to give the initial $O_G$ value. As $O_G$ is only relatively measured by OM contrast, the whole dataset is scaled between $0 < O_G < 1$. A justification of using $O_G$ for relative thickness measurements, using ellipsometry measurements, see SI Figure S5. The inhomogeneity of the oxide on each Cu orientation is defined as the coefficient of variation ($C_v$) of the relative oxidation post-normalisation with respect to the whole dataset:

$$C_v = \frac{\sigma}{\mu}$$

Where σ, the standard deviation, and μ, the mean, are both taken from the subset of data corresponding to a given crystallographic orientation of Cu. This measure was used to compensate for the significant differences in mean oxidation levels between Cu orientations and highlight inhomogeneity within those levels to enable comparison to other orientations.

**Scanning electron microscopy (SEM)**, with a Zeiss Gemini SEM, was used to map the $SiO_2$ substrate after transfer and the Cu tile after growth of graphene and etching. The samples were first levelled such that stage movements did not result in any change of focus of the substrates, then the manufacturer provided API was used to automate stage movement and take approximately 7000 images at a magnification of 600 x over 1x1cm$^2$. These images were then stitched and binarized to reveal areas where graphene was present, which could be used for the extraction of $T_G$.

**Close-spaced sublimation (CSS)** was done in a BM Pro 4" CVD reactor (base pressure 4 x 10-2 mbar) following previous work[15]. Single crystal MgO(168), single side polished (SurfaceNet), 1x1cm$^2$ crystals were rinsed in acetone then IPA (1 min each) before being dried in $N_2$ and loaded into the BM Pro. The MgO was placed 1 mm away from a planar polycrystalline Cu source (Alfa Aesar; 1mm thick; 99.9%). The source was then heated to 1075°C for 60 minutes, while the MgO substrate was approximately 950°C. This resulted in the epitaxial sublimation of Cu onto the MgO of the desired Cu orientation.

**Hall-bar devices** were fabricated with dry-transferred Gr, originating from Cu(168), and fully encapsulated by h-BN as in previous work.[9] The Hall-bar structure were defined in homogeneous regions with the lowest $G_{2D}$ with values around 16 cm$^{-1}$, indicating very small nanometer-scale strain variations of the graphene layer.[31] Further processed using electron beam lithography to define the shape, aluminium deposition to protect the region of interest, and Reactive Ion Etching with $SF_6$ to etch away the undesired material. A subsequent lithography step was performed to define contacts, and edge contacts were finally contacted with Cr/Au (5 nm/ 75 nm). For electrostatic gating, highly p-doped Si is used covered by a layer of 300 nm thick silicon oxide. The device geometry have a channel length of 4 μm and channel width of 3 μm. The h-BN / Gr / h-BN Hall-bar device sits on top of the silicon oxide layer and the h-BN has a thickness of roughly 20 nm.

**Electrical transport measurements** were performed on the Hall-bar devices in a vacuum pumped system. Standard lock-in techniques were used to measure the four-terminal resistance as well as Hall voltage and Hall conductivity. The charge carrier mobility μ as a function of charge carrier concentration n is calculated using the Drude formula $\sigma = ne\mu$, where $\sigma$ is the electrical conductivity. The electron mobility was extracted at a temperature of 300 K and a carrier concentration of $n = 1 \times 10^{12}$ $cm^{-2}$ to give $42.1 \times 10^3$ $cm^2$ $(Vs)^{-1}$.

## ASSOCIATED CONTENT

The following files are available free of charge.

Supplementary Information (.pdf) contains additional analysis and evidence to supplement claims in this manuscript.

## AUTHOR INFORMATION

Corresponding Author

Oliver J. Burton, ob303@cam.ac.uk

**Author Contributions**

The manuscript was written through contributions of all authors. All authors have given approval to the final version of the manuscript. ‡These authors were the main contributors to the research leading to this manuscript.

**Funding Sources**

We acknowledge funding from the EPSRC (EP/P005152/1, EP/T001038/1) and EU Horizon 2020 (grant agreement 785219). O.J.B. acknowledge an EPSRC Doctoral Training Award (EP/M508007/1).This project has received funding from the European Union's Horizon 2020 research and innovation program under grant agreement No. 881603 (Graphene Flagship) and from the European Research Council (ERC) under grant agreement No. 820254, the Deutsche Forschungsgemeinschaft (DFG, German Research Foundation) under Germany's Excellence Strategy - Cluster of Excellence Matter and Light for Quantum Computing (ML4Q) EXC 2004/1 - 390534769. K.W. and T.T. acknowledge support from JSPS KAKENHI (Grant Numbers 19H05790, 20H00354 and 21H05233).

## ABBREVIATIONS

CVD, Chemical Vapor Deposition; 2DM, 2D Material; h-BN, hexagonal boron nitride; IPF, inverse pole figure; PVA, polyvinyl alcohol; OM, optical microscope; SEM, scanning electron microscope; EBSD, electron backscatter diffraction; RCE, rotating compensator ellipsometry; IPA, isopropyl alcohol; XPS, X-ray photoemission spectroscopy; IE, imaging ellipsometry.

## REFERENCES

(1) Akinwande, D.; Huyghebaert, C.; Wang, C. H.; Serna, M. I.; Goossens, S.; Li, L. J.; Wong, H. S. P.; Koppens, F. H. L. Graphene and Two-Dimensional Materials for Silicon Technology. *Nature* **2019**, *573* (7775), 507–518. https://doi.org/10.1038/s41586-019-1573-9.

(2) Lemme, M. C.; Akinwande, D.; Huyghebaert, C.; Stampfer, C. 2D Materials for Future Heterogeneous Electronics. *Nat. Commun.* **2022**, *13* (1), 1392. https://doi.org/10.1038/s41467-022-29001-4.

(3) Huyghebaert, C.; Schram, T.; Smets, Q.; Kumar Agarwal, T.; Verreck, D.; Brems, S.; Phommahaxay, A.; Chiappe, D.; El Kazzi, S.; Lockhart de la Rosa, C.; Arutchelvan, G.; Cott, D.; Ludwig, J.; Gaur, A.; Sutar, S.; Leonhardt, A.; Marinov, D.; Lin, D.; Caymax, M.; Asselberghs, I.; et al. 2D Materials: Roadmap to CMOS Integration. In *2018 IEEE International Electron Devices Meeting (IEDM)*; 2018; p 22.1.1-22.1.4. https://doi.org/10.1109/IEDM.2018.8614679.

(4) Hofmann, S.; Braeuninger-Weimer, P.; Weatherup, R. S. CVD-Enabled Graphene Manufacture and Technology. *J. Phys. Chem. Lett.* **2015**, *6* (14), 2714–2721. https://doi.org/10.1021/acs.jpclett.5b01052.

(5) Deng, B.; Liu, Z.; Peng, H. Toward Mass Production of CVD Graphene Films. *Adv. Mater.* **2019**, *31* (9), 1800996. https://doi.org/10.1002/adma.201800996.

(6) Backes, C.; Abdelkader, A. M.; Alonso, C.; Andrieux-Ledier, A.; Arenal, R.; Azpeitia, J.; Balakrishnan, N.; Banszerus, L.; Barjon, J.; Bartali, R.; Bellani, S.; Berger, C.; Berger, R.; Ortega, M. M. B.; Bernard, C.; Beton, P. H.; Beyer, A.; Bianco, A.; Bøggild, P.; Bonaccorso, F.; et al. Production and Processing of Graphene and Related Materials. *2D Mater.* **2020**, *7* (2), 022001. https://doi.org/10.1088/2053-1583/ab1e0a.

(7) Fazio, D. D.; Purdie, D. G.; Ott, A. K.; Braeuninger-Weimer, P.; Khodkov, T.; Goossens, S.; Taniguchi, T.; Watanabe, K.; Livreri, P.; Koppens, F. H. L.; Hofmann, S.; Goykhman, I.; Ferrari, A. C.; Lombardo, A. High-Mobility, Wet-Transferred Graphene Grown by Chemical Vapor Deposition. *ACS Nano* **2019**, *13* (8), 8926–8935. https://doi.org/10.1021/acsnano.9b02621.

(8) Banszerus, L.; Schmitz, M.; Engels, S.; Dauber, J.; Oellers, M.; Haupt, F.; Watanabe, K.; Taniguchi, T.; Beschoten, B.; Stampfer, C. Ultrahigh-Mobility Graphene Devices from Chemical Vapor Deposition on Reusable Copper. *Sci. Adv.* **2015**, *1* (6). https://doi.org/10.1126/sciadv.1500222.

(9) Schmitz, M.; Ouaj, T.; Winter, Z.; Rubi, K.; Watanabe, K.; Taniguchi, T.; Zeitler, U.; Beschoten, B.; Stampfer, C. Fractional Quantum Hall Effect in CVD-Grown Graphene. *2D Mater.* **2020**, *7* (4), 041007. https://doi.org/10.1088/2053-1583/abae7b.

(10) Ullah, S.; Yang, X.; Ta, H. Q.; Hasan, M.; Bachmatiuk, A.; Tokarska, K.; Trzebicka, B.; Fu, L.; Rummeli, M. H. Graphene Transfer Methods: A Review. *Nano Res.* **2021**, *14* (11), 3756–3772. https://doi.org/10.1007/s12274-021-3345-8.

(11) Muñoz, R.; Gómez-Aleixandre, C. Review of CVD Synthesis of Graphene. *Chem. Vap. Depos.* **2013**, *19* (10–12), 297–322. https://doi.org/10.1002/cvde.201300051.

(12) Li, X.; Cai, W.; An, J.; Kim, S.; Nah, J.; Yang, D.; Piner, R.; Velamakanni, A.; Jung, I.; Tutuc, E.; Banerjee, S. K.; Colombo, L.; Ruoff, R. S. Large-Area Synthesis of High-Quality and Uniform Graphene Films on Copper Foils. *Science* **2009**, *324* (5932), 1312–1314. https://doi.org/10.1126/science.1171245.

(13) Sun, L.; Chen, B.; Wang, W.; Li, Y.; Zeng, X.; Liu, H.; Liang, Y.; Zhao, Z.; Cai, A.; Zhang, R.; Zhu, Y.; Wang, Y.; Song, Y.; Ding, Q.; Gao, X.; Peng, H.; Li, Z.; Lin, L.; Liu, Z. Toward Epitaxial Growth of Misorientation-Free Graphene on Cu(111) Foils. *ACS Nano* **2021**. https://doi.org/10.1021/acsnano.1c06285.

(14) Verguts, K.; Vermeulen, B.; Vrancken, N.; Schouteden, K.; Haesendonck, C. V.; Huyghebaert, C.; Heyns, M.; Gendt, S. D.; Brems, S. Epitaxial Al2O3(0001)/Cu(111) Template Development for CVD Graphene Growth. *J. Phys. Chem. C* **2016**, *120* (1), 297–304. https://doi.org/10.1021/acs.jpcc.5b09461.

(15) Burton, O. J.; Massabuau, F. C. P.; Veigang-Radulescu, V. P.; Brennan, B.; Pollard, A. J.; Hofmann, S. Integrated Wafer Scale Growth of Single Crystal Metal Films and High Quality Graphene. *ACS Nano* **2020**, *14* (10), 13593–13601. https://doi.org/10.1021/acsnano.0c05685.

(16) Luo, D.; Wang, M.; Li, Y.; Kim, C.; Yu, K. M.; Kim, Y.; Han, H.; Biswal, M.; Huang, M.; Kwon, Y.; Goo, M.; Camacho-Mojica, D. C.; Shi, H.; Yoo, W. J.; Altman, M. S.; Shin, H. J.; Ruoff, R. S. Adlayer-Free Large-Area Single Crystal Graphene Grown on a Cu(111) Foil. *Adv. Mater.* **2019**, *31* (35), 1903615. https://doi.org/10.1002/adma.201903615.

(17) Lu, A. Y.; Wei, S. Y.; Wu, C. Y.; Hernandez, Y.; Chen, T. Y.; Liu, T. H.; Pao, C. W.; Chen, F. R.; Li, L. J.; Juang, Z. Y. Decoupling of CVD Graphene by Controlled Oxidation of Recrystallized Cu. *RSC Adv.* **2012**, *2* (7), 3008–3013. https://doi.org/10.1039/c2ra01281b.

(18) Wang, R.; Whelan, P. R.; Braeuninger-Weimer, P.; Tappertzhofen, S.; Alexander-Webber, J. A.; Veldhoven, Z. A. V.; Kidambi, P. R.; Jessen, B. S.; Booth, T.; Bøggild, P.; Hofmann, S. Catalyst Interface Engineering for Improved 2D Film Lift-Off and Transfer. *ACS Appl. Mater. Interfaces* **2016**, *8* (48), 33072–33082. https://doi.org/10.1021/acsami.6b11685.

(19) Braeuninger-Weimer, P.; Burton, O. J.; Zeller, P.; Amati, M.; Gregoratti, L.; Weatherup, R. S.; Hofmann, S. Crystal Orientation Dependent Oxidation Modes at the Buried Graphene-Cu Interface. *Chem. Mater.* **2020**, *32* (18), 7766–7776. https://doi.org/10.1021/acs.chemmater.0c02296.

(20) Xu, X.; Yi, D.; Wang, Z.; Yu, J.; Zhang, Z.; Qiao, R.; Sun, Z.; Hu, Z.; Gao, P.; Peng, H.; Liu, Z.; Yu, D.; Wang, E.; Jiang, Y.; Ding, F.; Liu, K. Greatly Enhanced Anticorrosion of Cu by Commensurate Graphene Coating. *Adv. Mater.* **2018**, *30* (6), 1–7. https://doi.org/10.1002/adma.201702944.

(21) Luo, D.; Wang, X.; Li, B.-W.; Zhu, C.; Huang, M.; Qiu, L.; Wang, M.; Jin, S.; Kim, M.; Ding, F.; Ruoff, R. S. The Wet-Oxidation of a Cu(111) Foil Coated by Single Crystal Graphene. *Adv. Mater.* **2021**, *33* (37), 2102697. https://doi.org/10.1002/adma.202102697.

(22) Álvarez-Fraga, L.; Rubio-Zuazo, J.; Jiménez-Villacorta, F.; Climent-Pascual, E.; Ramírez-Jiménez, R.; Prieto, C.; Andrés, A. D. Oxidation Mechanisms of Copper under Graphene: The Role of Oxygen Encapsulation. *Chem. Mater.* **2017**, *29* (7), 3257–3264. https://doi.org/10.1021/acs.chemmater.7b00554.

(23) Luo, D.; You, X.; Li, B. W.; Chen, X.; Park, H. J.; Jung, M.; Ko, T. Y.; Wong, K.; Yousaf, M.; Chen, X.; Huang, M.; Lee, S. H.; Lee, Z.; Shin, H. J.; Ryu, S.; Kwak, S. K.; Park, N.; Bacsa, R. R.; Bacsa, W.; Ruoff, R. S. Role of Graphene in Water-Assisted Oxidation of Copper in Relation to Dry Transfer of Graphene. *Chem. Mater.* **2017**, *29* (10), 4546–4556. https://doi.org/10.1021/acs.chemmater.7b01276.

(24) Gao, L.; Guest, J. R.; Guisinger, N. P. Epitaxial Graphene on Cu(111). *Nano Lett.* **2010**, *10* (9), 3512–3516. https://doi.org/10.1021/nl1016706.

(25) Cançado, L. G.; Jorio, A.; Ferreira, E. H. M.; Stavale, F.; Achete, C. A.; Capaz, R. B.; Moutinho, M. V. O.; Lombardo, A.; Kulmala, T. S.; Ferrari, A. C. Quantifying Defects in Graphene via Raman Spectroscopy at Different Excitation Energies. *Nano Lett.* **2011**, *11* (8), 3190–3196. https://doi.org/10.1021/nl201432g.

(26) Lee, J. E.; Ahn, G.; Shim, J.; Lee, Y. S.; Ryu, S. Optical Separation of Mechanical Strain from Charge Doping in Graphene. *Nat. Commun.* **2012**, *3* (May), 1024. https://doi.org/10.1038/ncomms2022.

(27) Kim, M. S.; Kim, K. J.; Kim, M.; Lee, S.; Lee, K. H.; Kim, H.; Kim, H. M.; Kim, K. B. Cu Oxidation Kinetics through Graphene and Its Effect on the Electrical Properties of Graphene. *RSC Adv.* **2020**, *10* (59), 35671–35680. https://doi.org/10.1039/d0ra06301k.

(28) Calizo, I.; Bejenari, I.; Rahman, M.; Liu, G.; Balandin, A. A. Ultraviolet Raman Microscopy of Single and Multilayer Graphene. *J. Appl. Phys.* **2009**, *106* (4). https://doi.org/10.1063/1.3197065.

(29) Berciaud, S.; Li, X.; Htoon, H.; Brus, L. E.; Doorn, S. K.; Heinz, T. F. Intrinsic Line Shape of the Raman 2D-Mode in Freestanding Graphene Monolayers. *Nano Lett.* **2013**, *13* (8), 3517–3523. https://doi.org/10.1021/nl400917e.

(30) Frank, O.; Vejpravova, J.; Holy, V.; Kavan, L.; Kalbac, M. Interaction between Graphene and Copper Substrate: The Role of Lattice Orientation. *Carbon* **2014**, *68*, 440–451. https://doi.org/10.1016/j.carbon.2013.11.020.

(31) Neumann, C.; Reichardt, S.; Venezuela, P.; Drögeler, M.; Banszerus, L.; Schmitz, M.; Watanabe, K.; Taniguchi, T.; Mauri, F.; Beschoten, B.; Rotkin, S. V.; Stampfer, C. Raman Spectroscopy as Probe of Nanometre-Scale Strain Variations in Graphene. *Nat. Commun.* **2015**, *6* (May), 1–7. https://doi.org/10.1038/ncomms9429.

(32) Deng, Y.; Handoko, A. D.; Du, Y.; Xi, S.; Yeo, B. S. In Situ Raman Spectroscopy of Copper and Copper Oxide Surfaces during Electrochemical Oxygen Evolution Reaction: Identification of CuIII Oxides as Catalytically Active Species. *ACS Catal.* **2016**, *6* (4), 2473–2481. https://doi.org/10.1021/acscatal.6b00205.

(33) Luo, B.; Whelan, P. R.; Shivayogimath, A.; Mackenzie, D. M. A.; Bøggild, P.; Booth, T. J. Copper Oxidation through Nucleation Sites of Chemical Vapor Deposited Graphene. *Chem. Mater.* **2016**, *28* (11), 3789–3795. https://doi.org/10.1021/acs.chemmater.6b00752.

(34) Yu, Q.; Jauregui, L. A.; Wu, W.; Colby, R.; Tian, J.; Su, Z.; Cao, H.; Liu, Z.; Pandey, D.; Wei, D.; Chung, T. F.; Peng, P.; Guisinger, N. P.; Stach, E. A.; Bao, J.; Pei, S. S.; Chen, Y. P. Control and Characterization of Individual Grains and Grain Boundaries in Graphene Grown by Chemical Vapour Deposition. *Nat. Mater.* **2011**, *10* (6), 443–449. https://doi.org/10.1038/nmat3010.

(35) Nguyen, V. L.; Shin, B. G.; Duong, D. L.; Kim, S. T.; Perello, D.; Lim, Y. J.; Yuan, Q. H.; Ding, F.; Jeong, H. Y.; Shin, H. S.; Lee, S. M.; Chae, S. H.; Vu, Q. A.; Lee, S. H.; Lee, Y. H. Seamless Stitching of Graphene Domains on Polished Copper (111) Foil. *Adv. Mater.* **2015**, *27* (8), 1376–1382. https://doi.org/10.1002/adma.201404541.

(36) Yuan, Q.; Song, G.; Sun, D.; Ding, F. Formation of Graphene Grain Boundaries on Cu(100) Surface and a Route Towards Their Elimination in Chemical Vapor Deposition Growth. *Sci. Rep.* **2014**, *4* (1), 6541. https://doi.org/10.1038/srep06541.

(37) Brown, L.; Lochocki, E. B.; Avila, J.; Kim, C. J.; Ogawa, Y.; Havener, R. W.; Kim, D. K.; Monkman, E. J.; Shai, D. E.; Wei, H. I.; Levendorf, M. P.; Asensio, M.; Shen, K. M.; Park, J. Polycrystalline Graphene with Single Crystalline Electronic Structure. *Nano Lett.* **2014**, *14* (10), 5706–5711. https://doi.org/10.1021/nl502445j.

(38) Wu, R.; Ding, Y.; Yu, K. M.; Zhou, K.; Zhu, Z.; Ou, X.; Zhang, Q.; Zhuang, M.; Li, W.-D.; Xu, Z.; Altman, M. S.; Luo, Z. Edge-Epitaxial Growth of Graphene on Cu with a Hydrogen-Free Approach. *Chem. Mater.* **2019**, *31* (7), 2555–2562. https://doi.org/10.1021/acs.chemmater.9b00147.

(39) Rakheja, S.; Kumar, V.; Naeemi, A. Evaluation of the Potential Performance of Graphene Nanoribbons as On-Chip Interconnects. *Proc. IEEE* **2013**, *101* (7), 1740–1765. https://doi.org/10.1109/JPROC.2013.2260235.

(40) Couto, N. J. G.; Costanzo, D.; Engels, S.; Ki, D.-K.; Watanabe, K.; Taniguchi, T.; Stampfer, C.; Guinea, F.; Morpurgo, A. F. Random Strain Fluctuations as Dominant Disorder Source for High-Quality On-Substrate Graphene Devices. *Phys. Rev. X* **2014**, *4* (4), 041019. https://doi.org/10.1103/PhysRevX.4.041019.

(41) Banszerus, L.; Janssen, H.; Otto, M.; Epping, A.; Taniguchi, T.; Watanabe, K.; Beschoten, B.; Neumaier, D.; Stampfer, C. Identifying Suitable Substrates for High-Quality Graphene-Based Heterostructures. *2D Mater.* **2017**, *4* (2), 1–6. https://doi.org/10.1088/2053-1583/aa5b0f.

(42) Masubuchi, S.; Morimoto, M.; Morikawa, S.; Onodera, M.; Asakawa, Y.; Watanabe, K.; Taniguchi, T.; Machida, T. Autonomous Robotic Searching and Assembly of Two-Dimensional Crystals to Build van Der Waals Superlattices. *Nat. Commun.* **2018**, *9* (1), 1413. https://doi.org/10.1038/s41467-018-03723-w.

(43) Fan, Y.; Nakanishi, K.; Veigang-Radulescu, V. P.; Mizuta, R.; Stewart, J. C.; Swallow, J. E. N.; Dearle, A. E.; Burton, O. J.; Alexander-Webber, J. A.; Ferrer, P.; Held, G.; Brennan, B.; Pollard, A. J.; Weatherup, R. S.; Hofmann, S. Understanding Metal Organic Chemical Vapour Deposition of Monolayer WS<inf>2</Inf>: The Enhancing Role of Au Substrate for Simple Organosulfur Precursors. *Nanoscale* **2020**, *12* (43). https://doi.org/10.1039/d0nr06459a.

(44) Wang, S.; Dearle, A. E.; Maruyama, M.; Ogawa, Y.; Okada, S.; Hibino, H.; Taniyasu, Y. Catalyst-Selective Growth of Single-Orientation Hexagonal Boron Nitride toward High-Performance Atomically Thin Electric Barriers. *Adv. Mater.* **2019**, *31* (24). https://doi.org/10.1002/adma.201900880.

(45) Wang, R.; Purdie, D. G.; Fan, Y.; Massabuau, F. C. P.; Braeuninger-Weimer, P.; Burton, O. J.; Blume, R.; Schloegl, R.; Lombardo, A.; Weatherup, R. S.; Hofmann, S. A Peeling Approach for Integrated Manufacturing of Large Monolayer H-BN Crystals. *ACS Nano* **2019**, *13* (2), 2114–2126. https://doi.org/10.1021/acsnano.8b08712.

(46) Sutter, P.; Sadowski, J. T.; Sutter, E. A. Chemistry under Cover: Tuning Metal-Graphene Interaction by Reactive Intercalation. *J. Am. Chem. Soc.* **2010**, *132* (23), 8175–8179. https://doi.org/10.1021/ja102398n.

(47) Chin, H.-T.; Hofmann, M.; Huang, S.-Y.; Yao, S.-F.; Lee, J.-J.; Chen, C.-C.; Ting, C.-C.; Hsieh, Y.-P. Ultra-Thin 2D Transition Metal Monochalcogenide Crystals by Planarized Reactions. *Npj 2D Mater. Appl.* **2021**, *5* (1), 1–7. https://doi.org/10.1038/s41699-021-00207-2.

(48) Wang, L.; King, I.; Chen, P.; Bates, M.; Lunt, R. R. Epitaxial and Quasiepitaxial Growth of Halide Perovskites: New Routes to High End Optoelectronics. *APL Mater.* **2020**, *8* (10), 100904. https://doi.org/10.1063/5.0017172.

(49) Burton, O. J.; Babenko, V.; Veigang-Radulescu, V. P.; Brennan, B.; Pollard, A. J.; Hofmann, S. The Role and Control of Residual Bulk Oxygen in the Catalytic Growth of 2D

Materials. *J. Phys. Chem. C* **2019**, *123* (26), 16257–16267. https://doi.org/10.1021/acs.jpcc.9b03808.

(50) Costa, S. D.; Righi, A.; Fantini, C.; Hao, Y.; Magnuson, C.; Colombo, L.; Ruoff, R. S.; Pimenta, M. A. Resonant Raman Spectroscopy of Graphene Grown on Copper Substrates. *Solid State Commun.* **2012**, *152* (15), 1317–1320. https://doi.org/10.1016/j.ssc.2012.05.001.

# Supplementary Information: Putting high-index Cu on the map for high-yield, dry-transferred CVD graphene

*Oliver J. Burton[1‡*], Zachary C. M. Winter[2‡*], Kenji Watanabe[3], Takashi Taniguchi[4], Bernd Beschoten[2], Christoph Stampfer[2,5], Stephan Hofmann[1]*

[1]Department of Engineering, University of Cambridge, Cambridge CB3 0FA, United Kingdom

[2]2nd Institute of Physics A and JARA-FIT, RWTH Aachen University, 52074 Aachen, Germany

[3]Research Center for Functional Materials, National Institute for Materials Science, 1-1 Namiki Tsukuba, Ibaraki 305-0044, Japan

[4]International Center for Materials Nanoarchitectonics, National Institute for Materials Science, 1-1 Namiki Tsukuba, Ibaraki 305-0044, Japan

[5]Peter Grünberg Institute (PGI-9), Forschungszentrum Jülich, 52425 Jülich, Germany

[1] EBSD map stitching

[2]Oxidation apparatus/methodology

[3]Relative oxidation of Cu under graphene and Additional Raman data.

[4]Graphene orientation mapping with Cu orientation

[5]OM RCE correlation

[6] 2D widths on $SiO_2$

[7] XPS of Interfacial Oxide

[8] Schematic of tile processing for the high throughput study

[9] AFM before and after oxidation

[10] Raman sampling on copper before and after oxidation

[11] Method for extracting Raman data averages

[12] Dry-transfer process

[13] Comparison of Oxidation under graphene and on uncovered Cu

[14] Single crystal Cu graphene oxidation and transfer

## S1: EBSD Grain Detection and Stitching

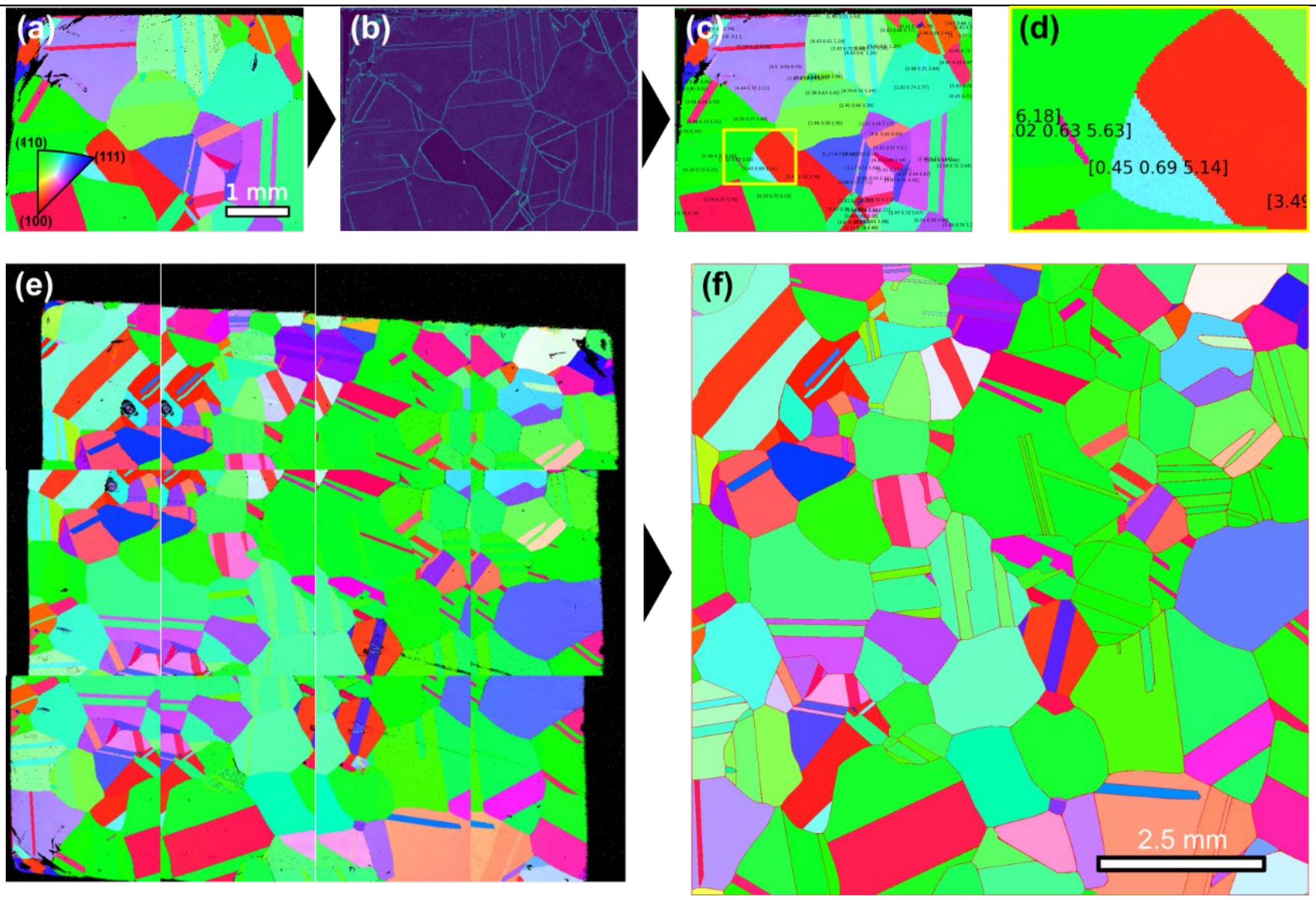

Figure S1: (a) the Z-IPF map of representing the raw data accumulated from the EBSD mapping. (b) The data represented in (a) after a mean difference operation has been used to show the presence of grain-boundaries and identify individual crystallographic orientations. (c) Directional mean Euler angles of the individual Cu grains as identified in (b). (d) The yellow region shown in (c) . (e) The unprocessed map consisting of 12 ill-aligned sub-maps. (f) The finished spatial orientation map (represented here as z-IPF colormap) used for this study.

**S2: Humidity-based Oxidation Chamber Schematic**

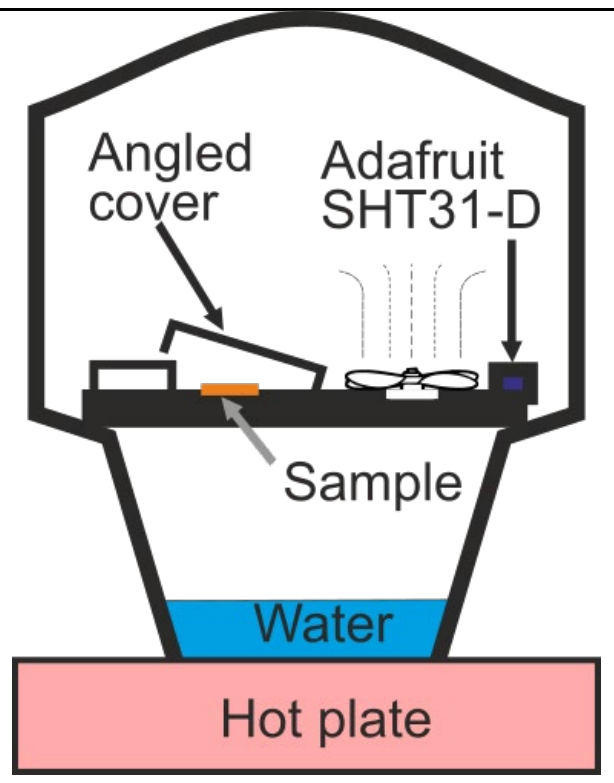

Figure S2: Schematic representation of the humidity-based oxidation apparatus and approximate sensor positioning. The oxidation was achieved by heating water in the base of closed desiccator on a hotplate to 70 °C, translating to a sample temperature of approximately 30 °C on the sample stage. A fan was placed on the sample stage to distribute water vapor and bring a stable internal humidity to > 99 %. Temperature and humidity were measured using an Adafruit SHT31-D connected to a Raspberry Pi. The entire set-up was placed in a temperature stabilized room to decrease internal temperature fluctuations. The sample was then covered with an angled glass plate to avoid any potential condensation falling onto the copper/graphene substrate.

## S3: Relative Oxidation Level ($O_G$) of Cu under Graphene

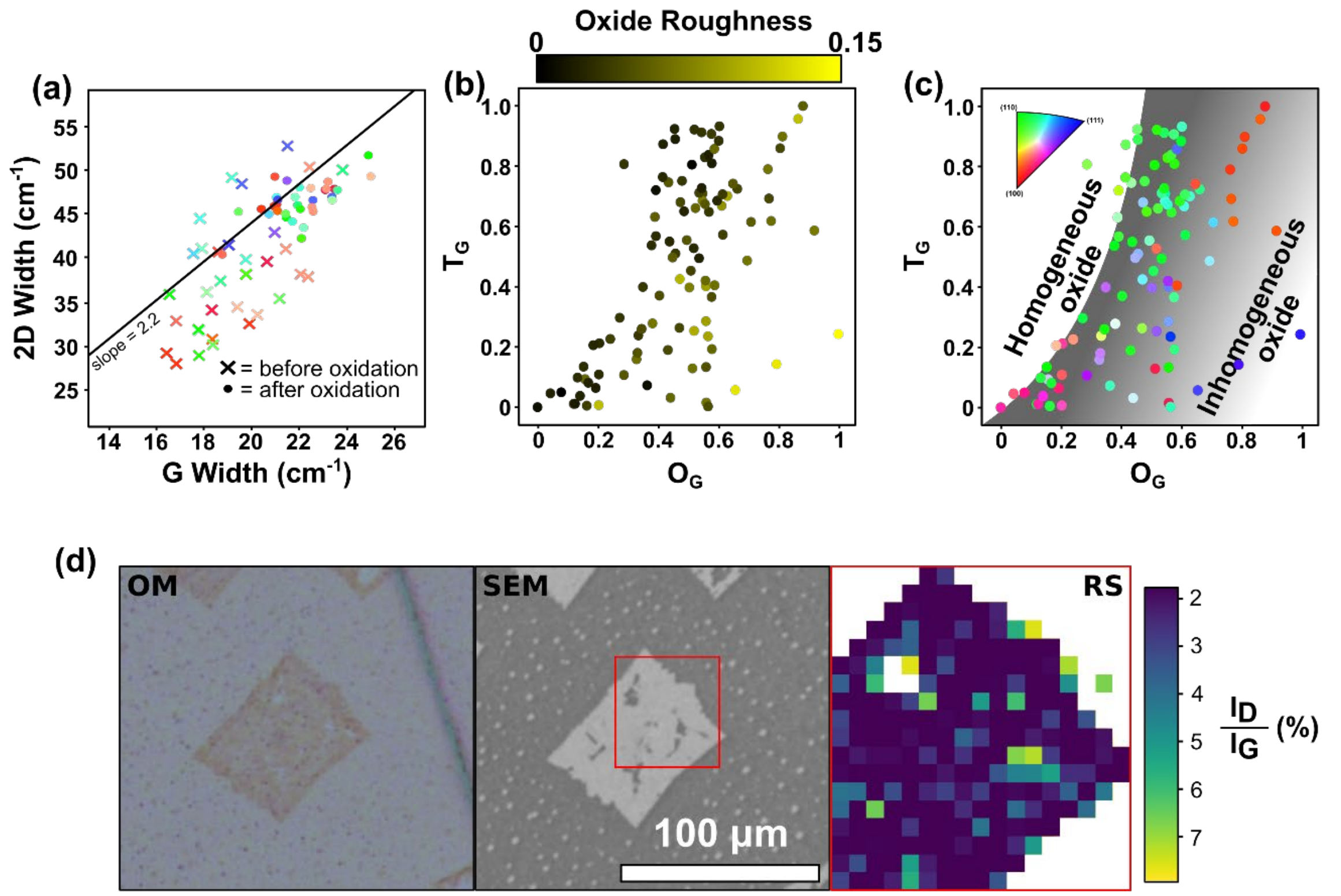

Figure S3: (a) The average Raman 2D peak width against the G peak width for graphene on Cu before (x) and after (o) oxidation. The color of this scatter plot is linked to the crystal orientation of the Cu that the graphene measurement was taken, according to the z-IPF inset in (c). (b) A reproduction of the data shown in Figure 2(d) to allow for comparison: a scatter plot of $T_G$ vs $O_G$ for all orientations measured, with the color of each point indicating the oxide roughness as defined in the methods. (c) The same data as in (b), where each point has been colored to indicate the crystallographic orientation to indicate some additional trends. (d) Example images of the island before (OM) and after (SEM) transfer, with a Raman map after transfer shown.

Figure S3 reinforces the data displayed in Figure 2 in the main text: with S3(c) showing some of the individual trends in oxidation with rough sections of the crystallographic space: towards

Cu(111) (indicated by blue) we note that these have a higher mean oxide thickness yet much lower transferred proportions, though the general trend of increasing oxide leading to increased transfer remains the same as in other orientations. Figure 3(d) highlights that the different $I_D/I_G$ of different Cu orientations is largely due to contributions from the cracks or areas where small amounts of graphene have not transferred and damaged the surrounding region.

## S4: Orientation Mapping of Gr on Cu

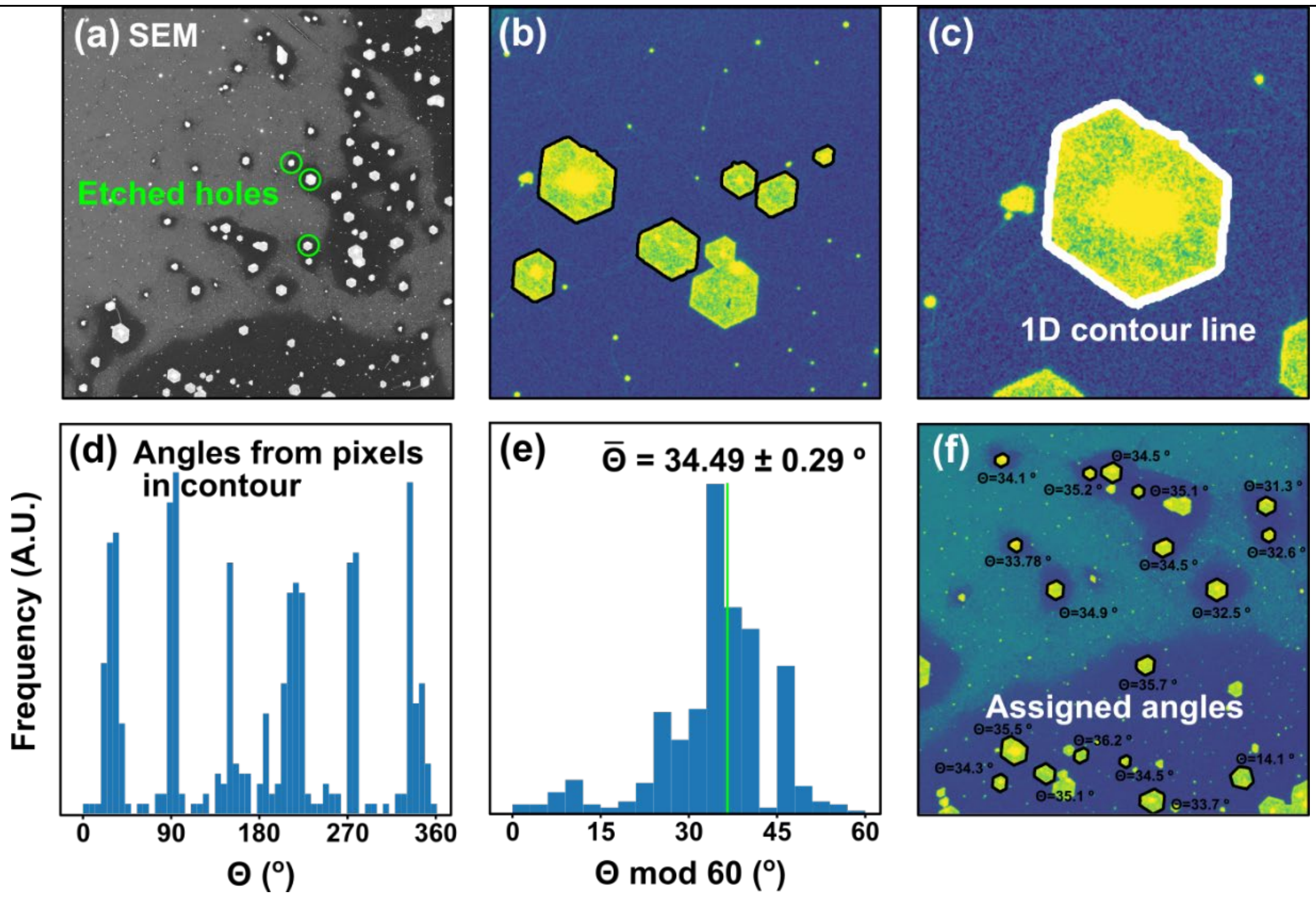

Figure S4-1: Island identification and principle angle determination. (a) SEM image of the etched holes in continuous graphene. (b) contrast-based contours with circularity > 0.7. (c) Example close-up image of etched hole contour line. (d) A histogram of the angles of points on the contour relative to neighboring (6 pixels along) points. (e) The angles from (d) condensed by a modulus function considering the constant angular offset of one side of a hexagon to another, allowing for the mean angle and standard error to be calculated using directional statistics. (f) A plot of the etched holes, showing which ones have been identified and the principal angles assigned to them.

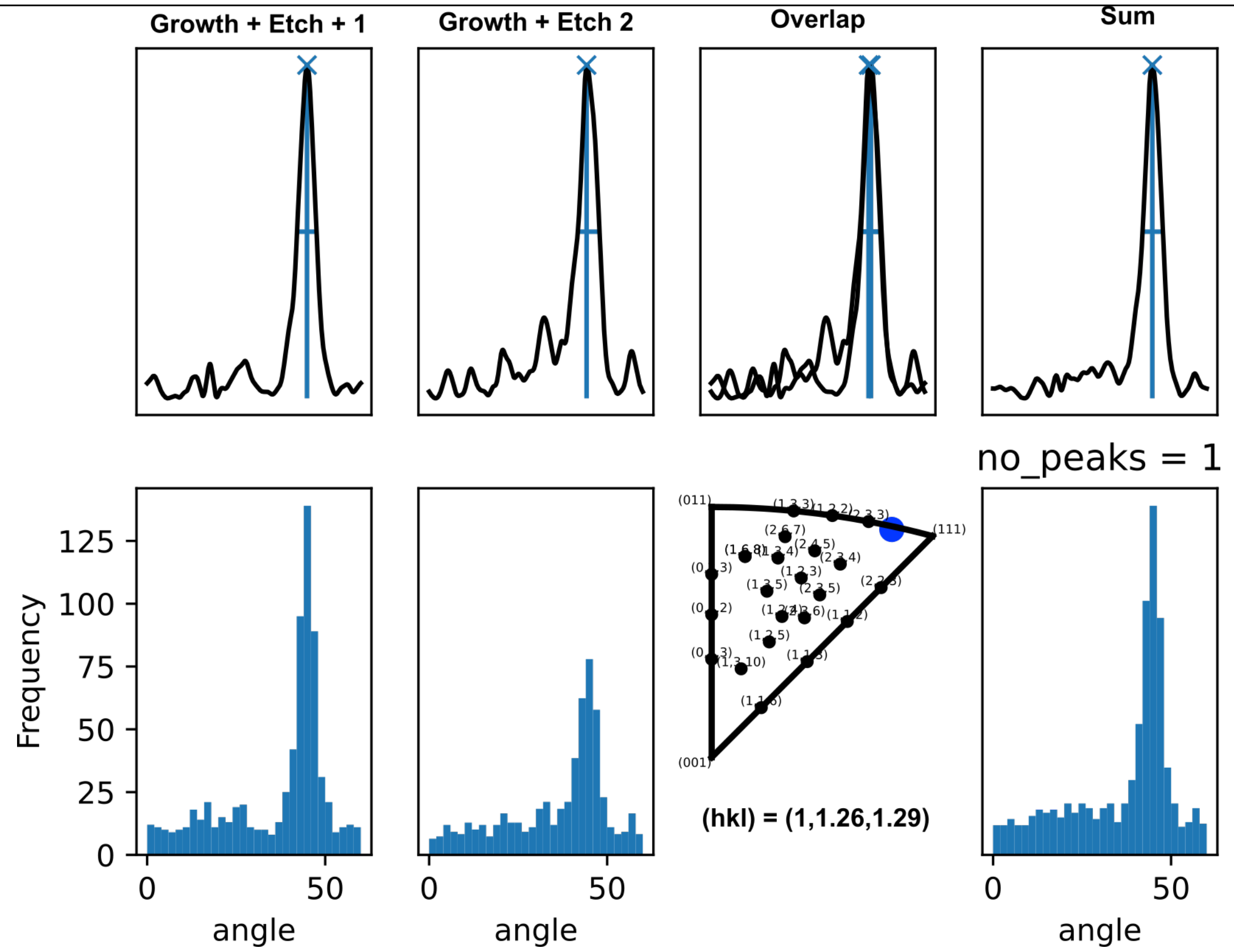

Figure S4-2: Example of the formation of a single segment of the plot on Figure 4: Two experimental datasets were created by the repeated growth of continuous film and subsequent etching detailed in previous literature[1], done on the same tile to both demonstrate experimental repeatability between growth and etches, and to increase signal strength. The resulting SEM image maps over 1x1 cm were processed using homebuilt code to extract the orientation of etched hexagonal domains. This data was then mapped on to the EBSD data to give each etched hexagonal domain detected an orientation and crystallographic orientation of the Cu. Etched holes were then grouped into datasets of orientation (i.e. each grain on the Cu), constructed into a histogram and associated density plot, where any peak(s) could be identified. As shown in the example case, the two independent datasets overwhelmingly correspond to each other, with each having one peak and the same position, and demonstrates

the repeatability of the orientation measurement technique. The final number of peaks identified is taken from the combination of both datasets and plotted in Figure 4.

To map the number of orientations of graphene on each Cu facet, a graphene etching process[1,2] was used to reveal the zig-zag edges of the graphene on the Cu, in the form of hexagonally etched holes, shown in Figure S4-1(a). Figure S4-1,2 shows the process of edge detection, inter-pixel measurement of that edge, binning and moduli to produce a histogram of measured angles (computationally). This was then analyzed with directional statistics to give a mean angle and associated standard deviation which was used to filter those etched holes or other sources of edges that were not hexagonal. Circularity and area were also used to minimize the effect of erroneous contributors to the statistics. The end result, after processing as detailed in the Methods section, was a histogram of etch hole angles that could be fitted with peaks to find the number of orientations present on a particular Cu orientataion.

**S5: Ellipsometry measurement of Oxide thickness**

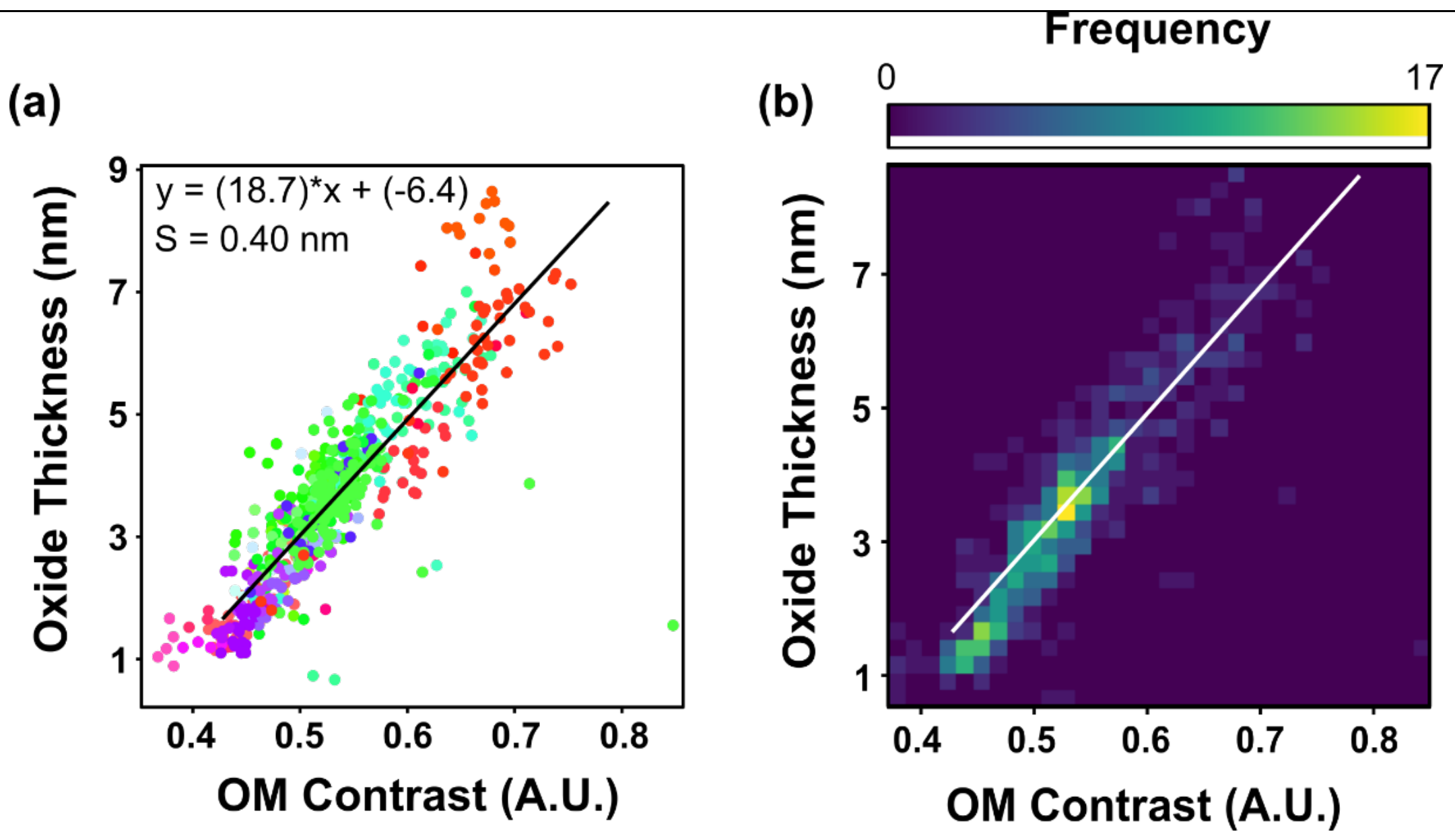

Figure S5: Correlation between OM contrast measurements and RCE determined Oxide thickness. (a) z-IPF colored scatter plot showing the relative OM contrast measurements plotted against RCE measurements of OM thickness. (b) the same as (a), where the plot is a density based heatmap where the color bar maps to the frequency of points in that bin. A least-squares linear fit has a correlation coefficient of 0.88, a strong correlation, and a standard error of 0.4 nm, illustrating that OM contrast is a reasonable method for relative oxide thickness measurement in the Cu system.

To demonstrate the correlation of OM images and oxide thickness measurements through rotating compensator ellipsometry (RCE), we take several local images, find the graphene domains in each image and then compare the average value within each graphene domain (i.e., each data point is the average of the values within a graphene domain). The Cu oxide thickness was fitted with Accurion's $Cu_2O$ model. This was done to compensate for a much lower spatial resolution given by the ellipsometry measurements. The results are summarized in Figure S5, showing a clear correlation and linear fit between the ellipsometry derived oxide thicknesses and the contrast derived relative thicknesses from OM. It is noted that when the Cu tile was mapped using the ellipsometers in-built software, there was not a linear spatial relationship between the ellipsometry data and the OM, SEM or RS measurements, thus this could not be used for the large-scale correlation as the other datatypes. It did however provide a range for the maximum and minimum levels of interfacial oxidation, between 0 and 10.2 nm, which given the linear and monotonic relationship shown here could be used in place of the relative oxidation measure ($O_G$) in Figures 2 and S3.

**S6: Raman Spectroscopy of Gr after Transfer, 2D and G peak widths.**

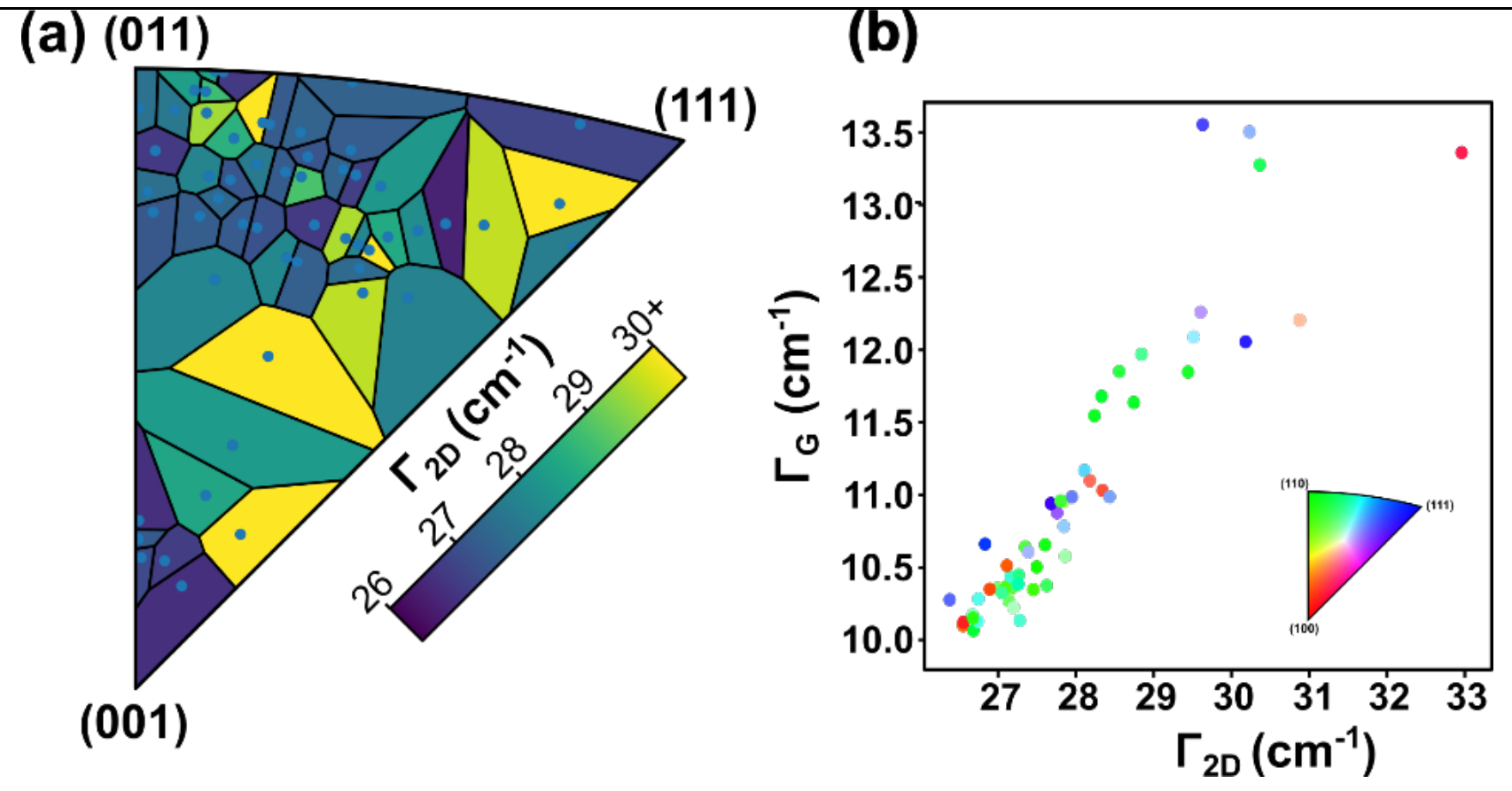

Figure S6: (a) The RS 2D peak width ($\Gamma_{2D}$) of graphene transferred onto $SiO_2$ with PVA as a function of crystallographic orientation of Cu. (b) A scatter plot of the $\Gamma_{2D}$ vs $\Gamma_G$, using the same data points as (a), where the color of each plotted point corresponds to the crystallographic orientation of the Cu the graphene was peeled from per inset IPF color reference.

Figure S6 shows the RS data plotted in an IPF map showing $\Gamma_{2D}$ as a function of the Cu orientation that the graphene was from and $\Gamma_{2D}$ vs $\Gamma_G$. $\Gamma_{2D}$ has been shown to correlate to the mobility of graphene-based field effect transistors[CITE], and as such could be used as a rough estimate for relative 'quality' of material along with the $I_D/I_G$ ratio shown in Figure 2.

## S7: XPS of Gr/Cu, Post Interfacial Oxide Formation

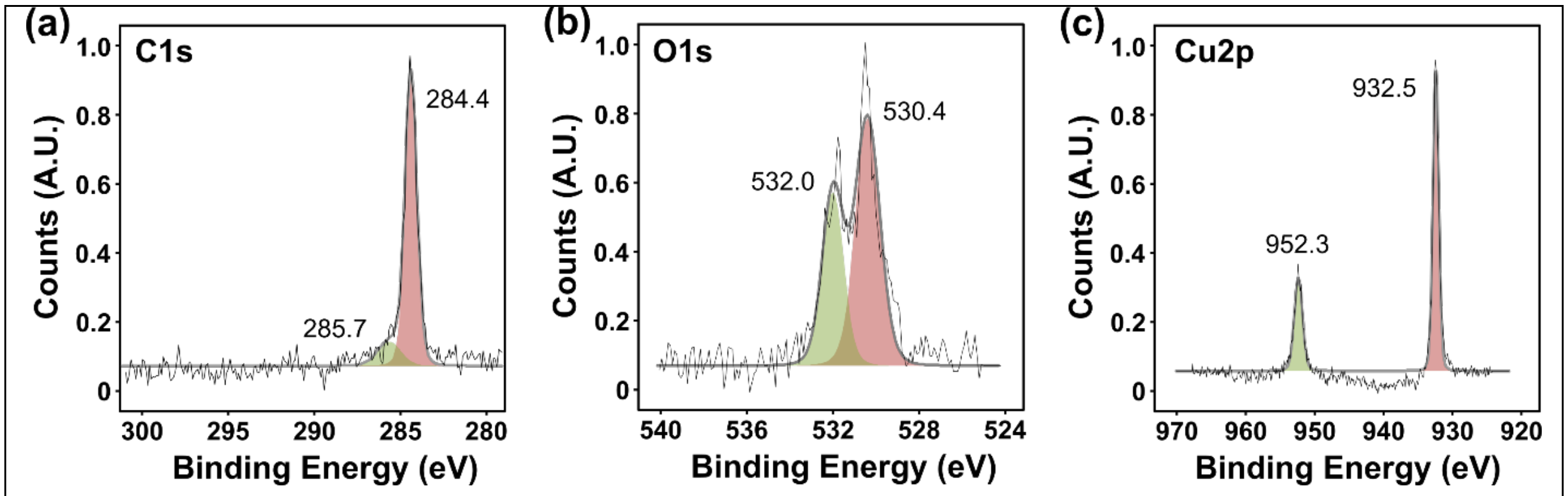

Figure S7: XPS measurements of graphene on Cu(168) after oxide formation with peak positions, determined by fitting the peaks with asymmetric Voight-like line profiles, shown adjacent to the highlighted peak. The black line indicates the experimental counts after background subtraction whilst the grey line indicates the sum of the individual fits. (a) The C1s peak with (red) carbon-carbon bonding and (green) carbon-oxygen bonding. (b) The O1s peak with (red) metal oxide and (green) metal carbonate sub-peaks. (c) The Cu2p peaks with (red) Cu2p3/2 and (green) Cu2p1/2 sub-peaks.

XPS measurements were conducted on Cu(111), Cu(120), Cu(121), Cu(122), Cu(123), Cu(236) and Cu(168) single crystals after graphene island growth and oxidation (See methods section). All samples showed near identical spectra, with graphene on Cu(168) after oxidation shown in Figure S7 as a representative example: no $Cu^{2+}$ satellites were observed (Figure S7(c)), despite copper oxide (Figure S7(b)) being present.[3] These results imply that the Cu oxide observed in this work is $Cu_2O$. Figure S7(a) shows the presence of graphitic or $Sp^2$ hybridized carbon (red), which we mainly attribute to the as grown graphene in this work; the shoulder (green) is attributed to carbon-oxygen bonding and is likely due to adventitious carbon contamination on the surface of the sample.[3]

## S8 Schematic Tile Process steps

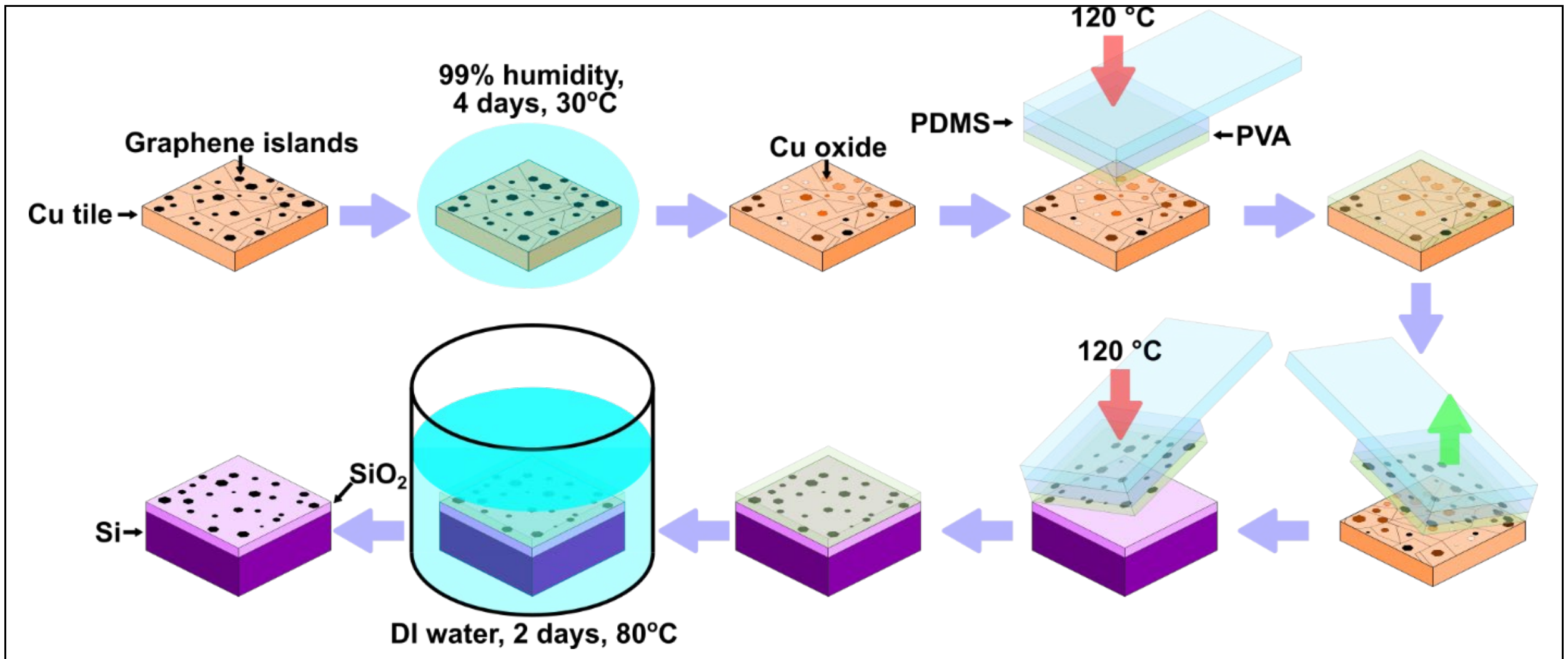

Figure S8: A Schematic of the process order for the growth of islands, interfacial oxidation, mechanical delamination and application to a target substrate.

### S9: AFM measurements of Cu and $Cu_2O$ regions on Cu(111)

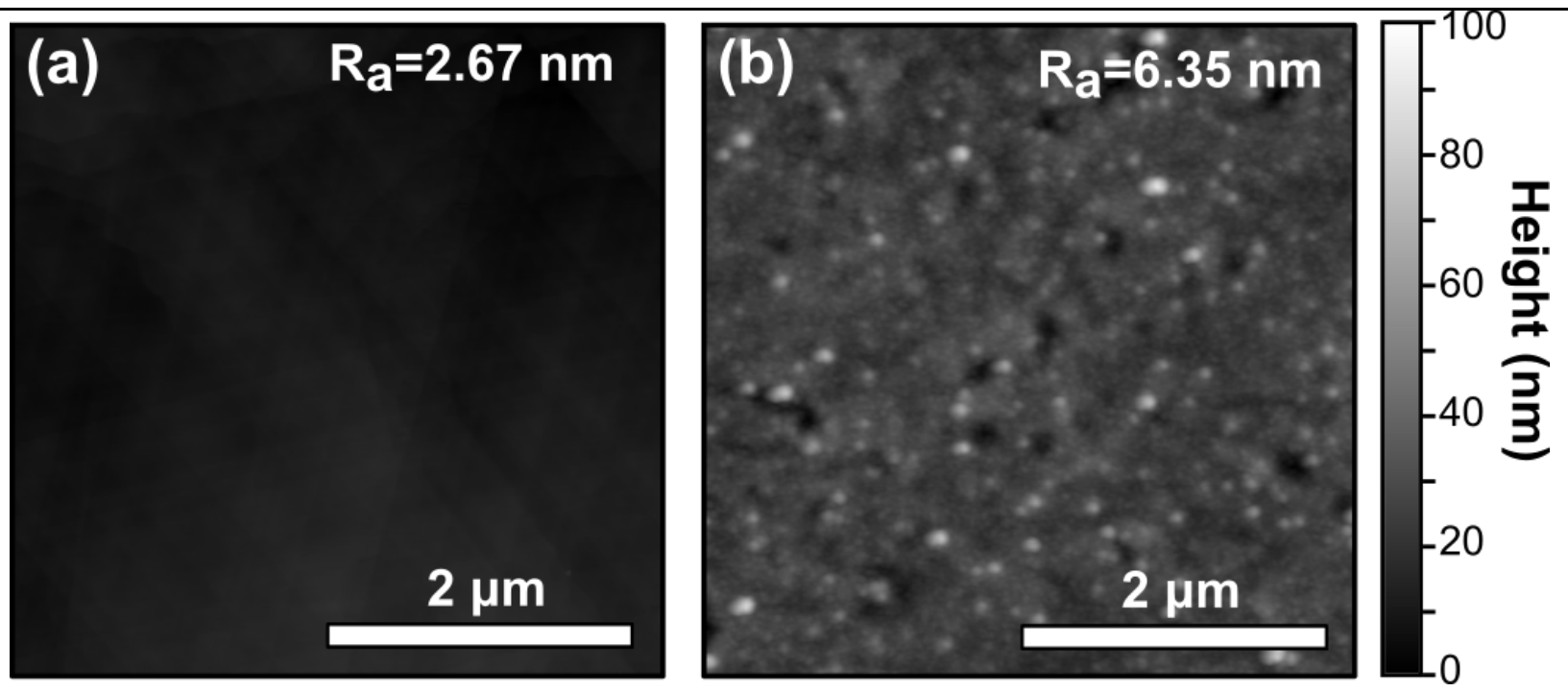

Figure S9: AFM maps before (a) and after (b) a typical oxidation process on a single Crystal Cu Sample, showing the increased surface roughness beneath the graphene as a result of the oxidation process.

## S10: Raman sampling on copper before and after oxidation

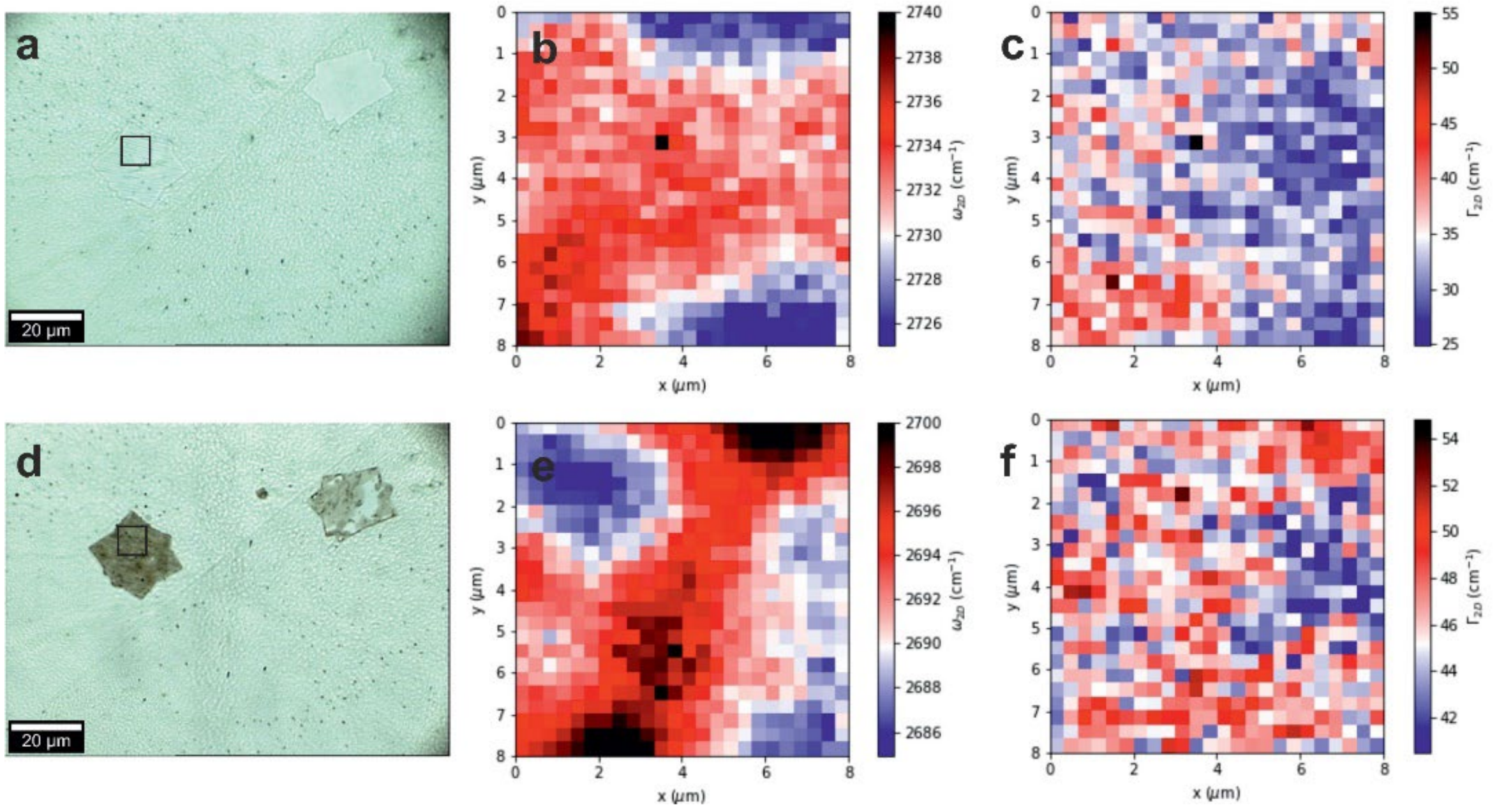

Figure S10: Example of Raman sampling for islands before (a-c) and after 4 days of oxidation on copper (d-f). Optical images of graphene islands on cooper are shown before and after oxidation in (a) and (d) and an 8x8 µm$^2$ area shows where Raman maps were taken. The 2D positions (b and d) and FWHM maps (c and d) are compared. The maps are subsequently processed as shown further in Figure S11.

## S11: Method for extracting Raman data averages

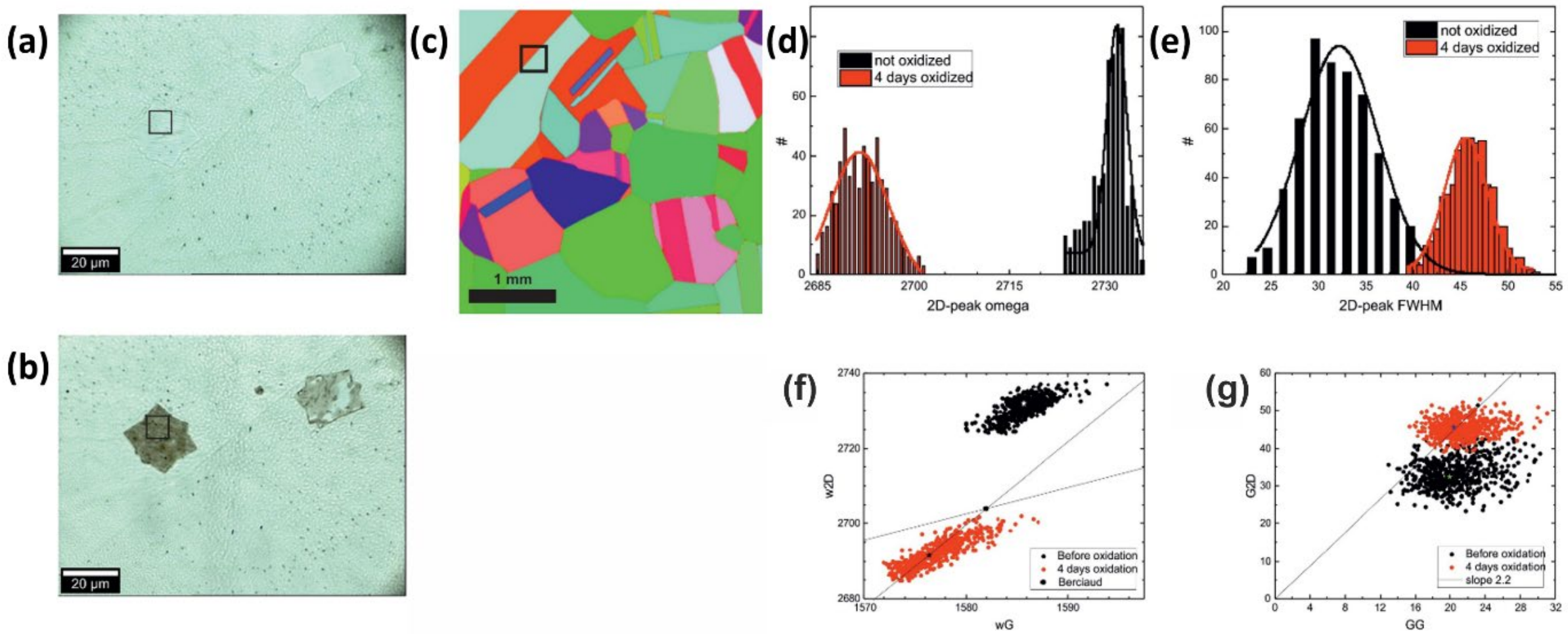

Figure S11 (a-b) show the graphene islands before and after oxidation. The copper orientation is known from EBSD measurements as shown in (c). The 8x8 μm2 Raman maps are condensed into histograms in (d-e) and the scatter plots are shown in (g-h). The average positions are plotted in Figure 2 (d) and widths in Figure S3.

## S12: Dry-transfer process

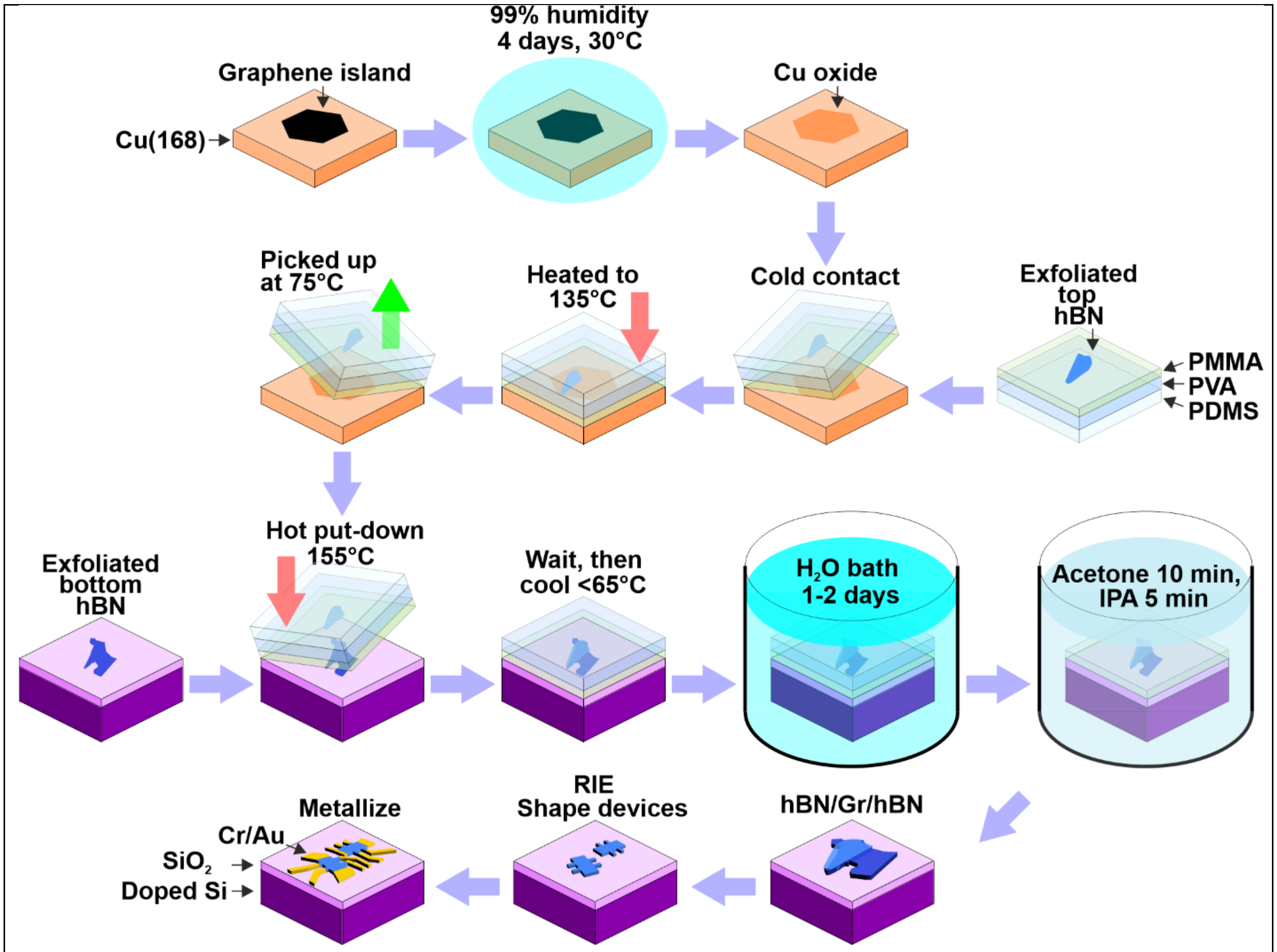

Figure S12: A schematic of the entire dry-transfer process from epitaxial Cu(168) substrate to finished Hall bar device. A single graphene island is picked-up with an exfoliated h-BN flake and placed down onto an exfoliated h-BN already on the target SiO2 substrate using the hot put-down technique. The polymers are removed and devices are structured using E-beam lithography, RIE for etching, and finally side-contacts are made using Cr/Au.

## S13: Comparison of Oxidation under graphene and on uncovered Cu

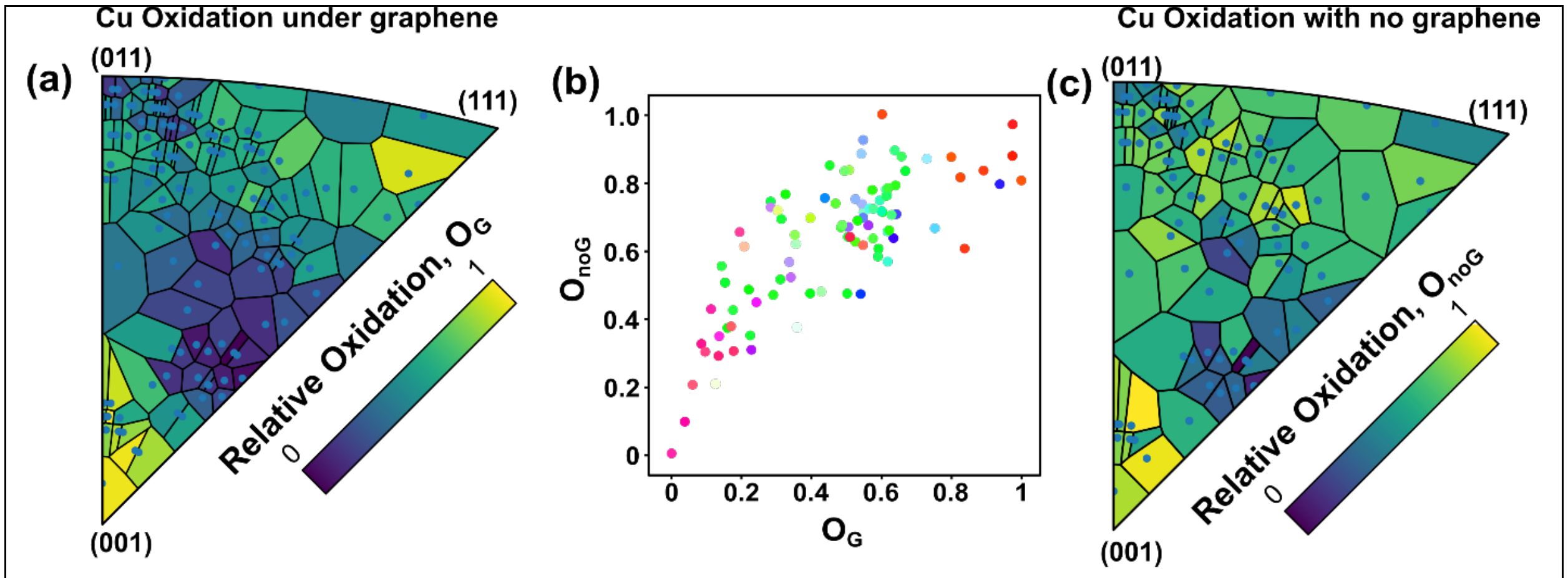

Figure S13: (a) A z-IPF showing the relative oxidation (defined in methods) of the Cu-Gr interface, $O_G$, dependent on crystallographic Cu orientation. (b) The relative oxidation of Cu facets on uncovered Cu against that of the graphene-Cu interface. This scatter plot effectively demonstrates a correlation between the oxidation of Cu uncovered and covered by graphene, with a Pearson correlation coefficient of 0.88. (c) A z-IPF showing the relative oxidation (defined in methods) of the uncovered Cu surface, $O_{noG}$, dependent on crystallographic Cu orientation.

## S14: Single crystal Cu graphene oxidation and transfer

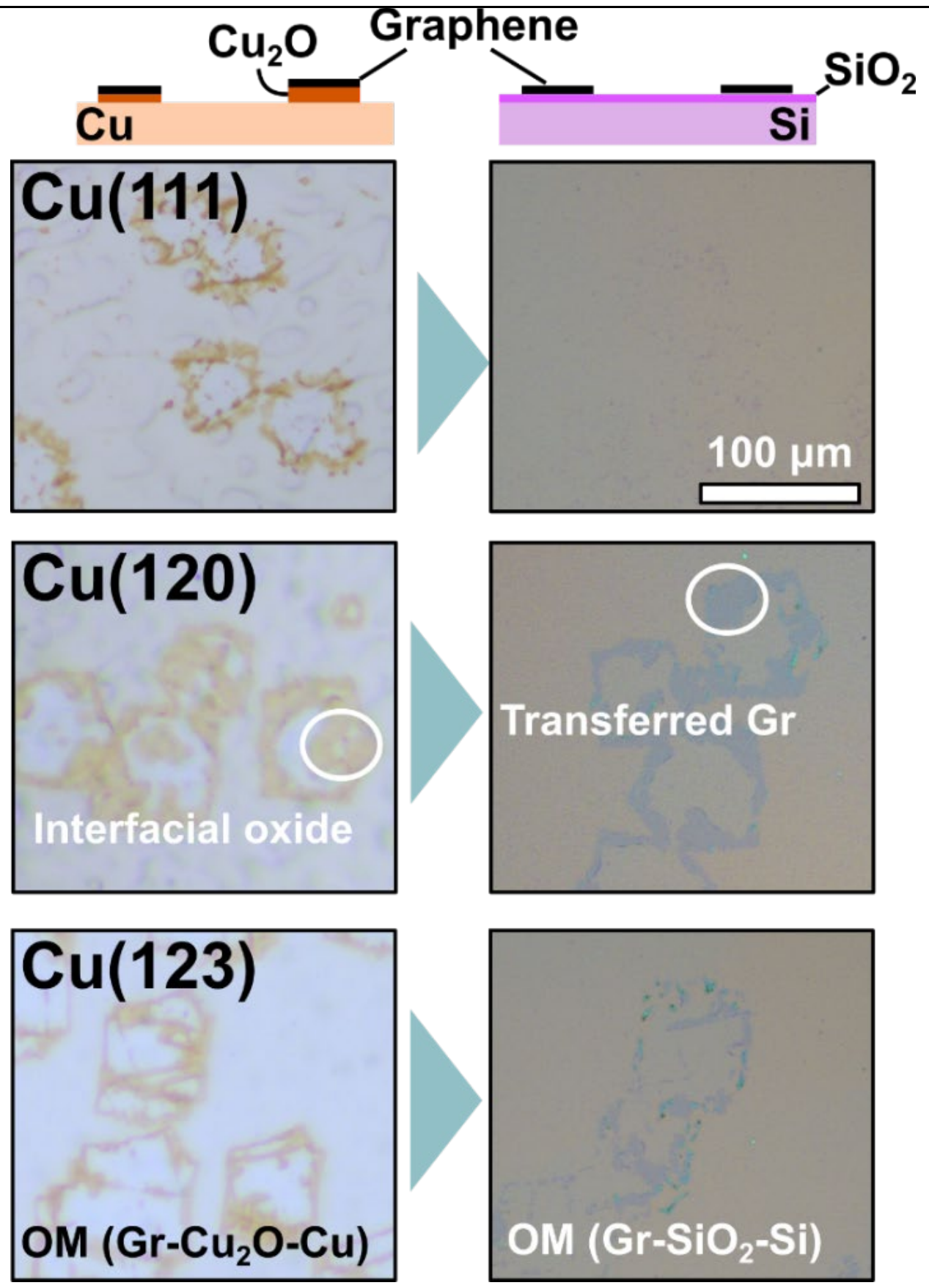

Figure S14: The oxidation of CSS deposited single crystal Cu samples with graphene grown and transferred, as per methods section. Similarities between the oxidization pattern and the transferred graphene can be clearly seen in Cu(120) and Cu(123), and again highlights the need for a consistent, homogeneous oxide. Cu(111) had only small flakes of graphene transferred, likely corresponding to the edge regions. Scale bar the same for all images.

**References**

1. Burton, O. J. *et al.* The Role and Control of Residual Bulk Oxygen in the Catalytic Growth of 2D Materials. *Journal of Physical Chemistry C* **123**, 16257–16267 (2019).
2. Burton, O. J. *et al.* Integrated wafer scale growth of single crystal metal films and high quality graphene. *ACS Nano* **14**, 13593–13601 (2020).
3. Moulder, J. F., Stickle, W. F., Sobol, P. E. & Bomben, K. D. *Handbook of X-ray photoelectron spectroscopy: a reference book of standard spectra for identification and interpretation of XPS data*. *Surface and Interface Analysis* (Perkin-Elmer Corporation, 1992).